\documentclass[sigconf, nonacm]{acmart}
\AtBeginDocument{%
  }

\usepackage{subfig}
\usepackage{amsmath}
\usepackage{algorithm2e}
\usepackage{algorithmic}
\usepackage{threeparttable}
\usepackage{enumitem}
\usepackage{booktabs}    % for \toprule, \midrule, etc.
\usepackage{multirow}    % for \multirow
\usepackage{graphicx}    % for \resizebox
\usepackage{siunitx}     % for S column type
\begin{document}

\title{Turning Semantics into Topology: LLM-Driven Attribute Augmentation for Collaborative Filtering}

\author{Junjie Meng}
\orcid{0009-0006-5709-9963}
\affiliation{
  \institution{School of Artificial Intelligence and Data Science, University of Science and Technology of China}
  \city{Hefei}
  \country{China}
}
\email{mengjre@gmail.com}

\author{Ranxu Zhang}
\orcid{0009-0006-7222-2149}
\affiliation{
  \institution{School of Artificial Intelligence and Data Science, University of Science and Technology of China}
  \city{Hefei}
  \country{China}
}
\email{zkd_zrx@mail.ustc.edu.cn}

\author{Wei Wu}
% \authornote{Chao Wang is the corresponding author.}
% \orcid{0000-0001-7717-447X}
\affiliation{
  \institution{School of Artificial Intelligence and Data Science, University of Science and Technology of China}
  % \department{}
  \city{Hefei}
  \country{China}
}
\email{urara@mail.ustc.edu.cn}

\author{Rui Zhang}
% \authornote{Chao Wang is the corresponding author.}
% \orcid{0000-0001-7717-447X}
\affiliation{
  \institution{Alibaba Group}
  % \department{}
  \city{Hangzhou}
  \country{China}
}
\email{modun.zr@taobao.com}

\author{Chuan Qin}
% \authornote{Chao Wang is the corresponding author.}
% \orcid{0000-0001-7717-447X}
\affiliation{
  \institution{Computer Network Information Center, Chinese Academy of Sciences}
  % \department{}
  \city{Beijing}
  \country{China}
}
\email{chuanqin0426@gmail.com}

\author{Zhang Qi}
% \authornote{Chao Wang is the corresponding author.}
% \orcid{0000-0001-7717-447X}
\affiliation{
  \institution{Shanghai Artificial Intelligence Laboratory}
  % \department{}
  \city{Shanghai}
  \country{China}
}
\email{zhangqi.fqz@gmail.com}

\author{Qi Liu}
\affiliation{
  \institution{School of Artificial Intelligence and Data Science, University of Science and Technology of China}
  \city{Hefei}
  \country{China}
}
\email{qiliuql@ustc.edu.cn}

\author{Hui Xiong}
\orcid{0000-0001-6016-6465}
\affiliation{%
  \institution{Thrust of Artificial Intelligence, The Hong Kong University of Science and Technology (Guangzhou)}
  \city{Guangzhou}
  \country{China}
  \institution{Department of Computer Science and Engineering,The Hong Kong University of Science and Technology}
  \city{Hong Kong SAR}
  \country{China}
}

\email{xionghui@ust.hk}

\author{Chao Wang}
% \authornote{Chao Wang is the corresponding author.}
\orcid{0000-0001-7717-447X}
\affiliation{
  \institution{School of Artificial Intelligence and Data Science, University of Science and Technology of China}
  \city{Hefei}
  \country{China}
}
\email{wangchaoai@ustc.edu.cn}
\begin{abstract}
Large Language Models (LLMs) have shown great potential for enhancing recommender systems through their extensive world knowledge and reasoning capabilities. However, effectively translating these semantic signals into traditional collaborative embeddings remains an open challenge. Existing approaches typically fall into two extremes: direct inference methods are computationally prohibitive for large-scale retrieval, while embedding-based methods primarily focus on unilateral feature augmentation rather than holistic collaborative signal enhancement. To bridge this gap, we propose \textbf{\underline{T}}opology-\textbf{\underline{A}}ugmented \textbf{\underline{G}}raph \textbf{\underline{C}}ollaborative \textbf{\underline{F}}iltering (TAGCF), a novel framework that transforms semantic knowledge into topological connectivity. Unlike existing approaches that depend on textual features or direct interaction synthesis, TAGCF employs LLMs to infer interaction intents and underlying causal relationships from user-item pairs, representing these insights as intermediate attribute nodes within an enriched User-Attribute-Item (U-A-I) graph. Furthermore, to effectively model the heterogeneous relations in this augmented structure, we propose \textbf{\underline{A}}daptive \textbf{\underline{R}}elation-weighted \textbf{\underline{G}}raph \textbf{\underline{C}}onvolution (ARGC), which employs relation-specific prediction networks to dynamically estimate the importance of each relation type. Extensive experiments across multiple benchmark datasets and CF backbones demonstrate consistent improvements, with comprehensive evaluations including cold-start scenarios validating the effectiveness and robustness of our framework. All code will be made publicly available. For anonymous review, our code is available at the following anonymous link: https://anonymous.4open.science/r/AGCF-2441353190/.
\end{abstract}



\maketitle

\section{Introduction}

Collaborative filtering (CF) based recommendation methods~\cite{10.1561/1100000009CFRS,koren2009mf,he2017neural,wang2022profiling,pan2022collaborative} have achieved remarkable success by leveraging historical user-item interactions to predict user preferences. However, the inherent sparsity of real-world interaction data poses significant challenges, especially in the cold-start scenarios. To address this limitation, content-based~\cite{yu2018aesthetic,chen_ACF} and knowledge graph-based~\cite{wang2019kgat,wang2020knowledge,xia2021knowledge} CF approaches incorporate auxiliary knowledge such as item descriptions and entity relationships. Nevertheless, acquiring such additional knowledge often requires substantial human and computational resources.
Recently, the rapid advancement of Large Language Models (LLMs)~\cite{wu2024surveylargelanguagemodels,zhao2025surveylargelanguagemodels,zhang2025m2recmultiscalemambaefficient,10.1007/978-3-031-56060-6_24llmzs} has attracted significant interest from the recommender systems community~\cite{wang-etal-2023-rethinking-evaluation,10411560,wang-etal-2023-chatgpt}. LLMs offer a promising paradigm for automated knowledge enhancement by providing semantic understanding and world knowledge that can enrich sparse interactions without requiring extensive manual curation.

\begin{figure*}[h!]
    \centering
    \includegraphics[width=\linewidth]{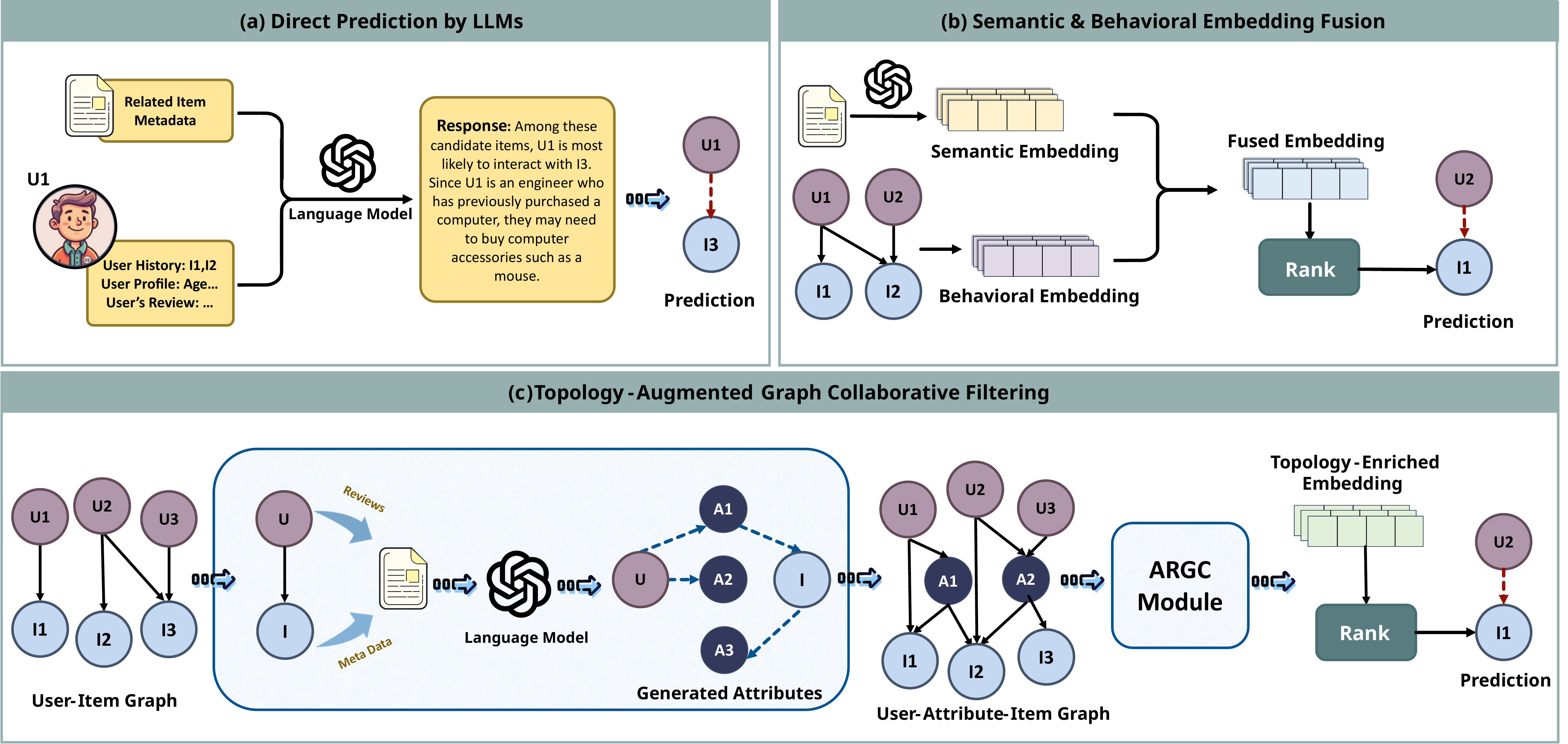}
    % \captionsetup{aboveskip=4pt, belowskip=2pt}
    \caption{Comparison with two main LLM-based collaborative methods with TAGCF.}
    \label{fig:compare}
\end{figure*}

As shown in Figure~\ref{fig:compare}a and Figure~\ref{fig:compare}b, LLMs have been integrated into recommender systems mainly in two ways: either by directly generating predictions or by providing embeddings for downstream models. 
On the one hand, a primary focus involves employing LLMs as the final predictor. For instance, LLMRec~\cite{10.1145/3616855.3635853LLMRec} uses LLMs to generate possible interaction pairs to enrich user-item graph. RPP~\cite{10.1145/3716320RPP} applies reinforcement learning to search for user-specific prompts, thereby achieving more accurate LLM-based predictions. TallRec~\cite{bao2023tallrec} finetunes LLMs specifically on recommendation tasks and then leverages the models to output the final prediction. 
On the other hand, another significant stream of work focuses on utilizing pre-trained language models to extract semantic features from data as item descriptions or user profiles, which then serve to augment traditional CF models. 
RLMRec~\cite{Ren_2024RLM} employs LLMs to generate textual embeddings for users and items, aiming to align language and behavioral spaces through representation learning. 
AlphaRec~\cite{sheng2025language} initializes user and item representations with textual embeddings derived from pretrained LLMs, demonstrating that even after a simple linear mapping and graph convolution, such embeddings can outperform traditional ID-based methods. 

Despite their potential, LLM-based approaches face several critical challenges.
Direct LLM-based recommendation approaches leverage the reasoning capabilities of large models to make predictions, but these approaches are highly inefficient, typically limited to inferring over sampled subsets of items for a user due to context window constraints and inference costs.
Another mainstream direction, LLM-enhanced approaches fuse semantic embeddings and behavioral signals to rank all items in a single inference. However, these methods suffer from fundamental limitations in their integration with collaborative filtering. They primarily focus on unilateral information enhancement from either user or item perspectives, essentially expanding content-based features rather than leveraging LLMs' world knowledge about collaborative filtering similarities, causal relationships, and interaction intents inherent in user-item dynamics. More critically, the enhanced textual embeddings are not inherently aligned with the specific objectives of recommendation tasks. This misalignment becomes particularly pronounced in data sparsity and cold-start scenarios, where reliance on mismatched semantic features may fail to compensate for the lack of interaction signals. Therefore, it remains a critical challenge to leverage LLMs to extract latent structural signals inherent to the collaborative filtering perspective.

Drawing from both LLM-based interaction synthesis approaches~\cite{zhang2024textlikeencodingcollaborativeinformation,liao2025patchrecmultigrainedpatchingefficient,10882951CoLLM,DBLP:journals/corr/abs-2303-14524} and recent advances in auxiliary graph information utilization (e.g., knowledge graphs~\cite{park2022reinforcement,wang2022multi,wang2021learning} and social networks~\cite{wu2018social_collaborative,sun2018attentive_social,yu2020ESRF}), we derive two key insights that guide our framework design:
\textbf{Insight $\mathbf{1}$} \textit{The structural connectivity of user-item interaction networks is the primary driver of collaborative filtering performance, particularly for sparse nodes.}
\textbf{Insight $\mathbf{2}$} \textit{Semantic information does not necessarily require semantic embeddings.} 
 These motivate our design of LLM-extracted conceptual nodes as {intermediaries}—capturing the underlying motivations behind interactions—to achieve principled integration of extra information without using semantic embeddings.

To this end, we propose \textbf{T}opology-\textbf{A}ugmented \textbf{G}raph \textbf{C}ollaborative \textbf{F}iltering (\textbf{TAGCF}), a general framework that addresses these challenges by leveraging {offline} LLM-derived semantic knowledge through structural graph enhancement. Unlike existing approaches that rely on textual embeddings or direct interaction synthesis, TAGCF employs LLMs to infer interaction intents from user-item pairs, extracting these insights as intermediate attribute nodes in an enriched User-Attribute-Item (U-A-I) graph. Figure ~\ref{fig:compare}c displays this process.
% This design requires only single user-item queries per LLM call while creating semantically meaningful pathways that enhance collaborative signals throughout the recommendation pipeline.
To effectively handle the heterogeneous relations in the U-A-I graph, we propose \textbf{A}daptive \textbf{R}elation-weighted \textbf{G}raph \textbf{C}onvolution (\textbf{ARGC}), which employs relation-specific prediction networks to dynamically estimate the importance of each relation type. This enables adaptive aggregation that accounts for the different roles of user-attribute and attribute-item connections while preserving collaborative filtering coherence. 
Extensive experiments across various CF backbones and benchmark datasets demonstrate consistent improvements, with our approach outperforming methods that rely on text embeddings while maintaining computational efficiency. Additional cold-start evaluations, ablation studies, and parameter analyses further validate the effectiveness and robustness of our framework. Our main contributions are as follows:

\begin{itemize}[nosep]

    \item We propose \textbf{TAGCF}, a model-agnostic framework that leverages LLM-extracted attributes to augment interaction graphs through structural enhancement, achieving superior performance without utilizing textual embeddings.
    
    \item We introduce \textbf{ARGC}, a relation-aware convolution mechanism that enables principled aggregation across heterogeneous relations in the U-A-I graph, facilitating holistic collaborative signal enhancement.
    
    \item We demonstrate our framework's effectiveness across multiple CF backbones and datasets, showing consistent improvements over conventional augmentation strategies and text-embedding methods, with comprehensive analysis on Anti-Over-Smoothing as well as other features.
\end{itemize}
\vspace{-6pt}

\section{Preliminaries and Related Work}

This section provides a brief overview of graph-based collaborative filtering, which serves as the backbone of our work, and subsequently summarize existing methods on recommendation systems that leverage data augmentation.

\subsection{Graph-based Collaborative Filtering}
 Collaborative filtering learns from sparse user-item interaction data and produces behavior-based representation embeddings denoted as $\mathcal{X}$. Recent advanced recommender systems leverage graph neural networks (GNNs)~\cite{chen2018fastgcn,Wang2019NeuralGraphCollaborative,li2019deepgcns,zhang2019heterogeneous,niepert2016learning,gao2018large,10.1109/TKDE.2023.3288135XSimGCL} to aggregate neighborhood information and capture complex high-order connectivity patterns in the user-item interaction graph~\cite{tintarev2007survey,sinha2002role,kunkel2019let,zhang2014explicit,yang2020hagerec,xie2021improving}. The core objective is to learn optimal user and item embeddings $\mathbf{X} = {\mathbf{X}_u, \mathbf{X}_i}$ that maximize the posterior likelihood given the observed positive interactions $\mathcal{E}^+$:

\vspace{-5mm}
\begin{equation}
\mathbf{X}^* = \arg\max_{\mathbf{X}} p(\mathbf{X} \mid \mathcal{E}^+).
\label{eq:arg,max}
\end{equation}

Here, $p(\mathbf{X} \mid \mathcal{E}^+)$ reflects how well the embeddings encode user-item relationships from the interaction set $\mathcal{E}^+$, enabling accurate prediction of user preferences. Finally, the possibility for a user-item pair $\langle u, i\rangle$ is computed via the score function of their embeddings: $\hat{y}_{u,i} = f(\mathbf{x}_u, \mathbf{x}_i)$.

\subsection{Recommendation with Data Augmentation}
 Since collaborative filtering methods struggle with cold-start and data sparsity problems, many recommendation systems~\cite{gallicchio2020fast,lai2020policy,zhou2020deepgnn} have sought to incorporate auxiliary information to enrich the representations derived from interaction data. Earlier approaches primarily rely on existing external data sources, such as knowledge graphs~\cite{wang2019kgat,wang2020knowledge,xia2021knowledge} and social networks~\cite{chen2021graph,cialdini2004social,mcpherson2001birds}.
 % For example, KGCN~\cite{Wang_2019KGCN} focuses on item-side enrichment by learning representations from knowledge graphs through an auxiliary network, which are then fused with embeddings obtained from user-item interactions. To address the noise and long-tail issues often present in knowledge graphs, KGCL~\cite{Yang_2022KGCL} proposes a contrastive learning framework that enhances item representations by suppressing irrelevant or noisy signals during knowledge aggregation. From the user-side perspective, SocialGCN~\cite{LIAO2022595Social} leverages social relationships to learn user semantic representations from user-user graphs and integrates them with interaction-based embeddings derived via LightGCN~\cite{He2020LightGCN}.However, due to data limitations, most of these external signals are restricted to user-user or item-item relationships, which may not directly align with or may even distort the informative patterns captured in user-item interactions.
 
 With the rapid advancement of Large Language Models (LLMs), recent studies have begun to leverage their generative and reasoning capabilities to enrich recommendation signals or even perform end-to-end prediction. 
A common paradigm is to utilize LLMs to generate user- or item-related textual descriptions, which are subsequently encoded into dense embeddings and fused with collaborative filtering representations:
\begin{equation}
    \mathbf{X}_{emb} = \mathrm{Emb}(\mathrm{LLM}(\mathcal{D}_{extra})),
    \vspace{-1.5mm}
\end{equation}
\begin{equation}
    \vspace{-1mm}
    \mathbf{X} = \mathrm{Fusion}(\mathbf{X}_{cf}, \mathbf{X}_{emb}),
\end{equation}
where $\mathcal{D}_{extra}$ denotes the additional textual data generated or refined by the LLM, and $\mathbf{X}_{cf}$ represents the embeddings learned from user–item interaction graph. The fusion operation can be realized via concatenation, linear transformation\cite{sheng2025language}, or attention-based aggregation.

In contrast, another research line directly employs LLMs as the final recommender or scoring function, where the model predicts user preference through text-based inferring:
\begin{equation}
    \hat{y}_{u,i} = \mathrm{LLM}(\mathrm{Prompt}(u, i, \mathcal{H}_u, \mathcal{H}_i)),
\end{equation}
where $\mathcal{H}_u$ and $\mathcal{H}_i$ denote the historical interaction contexts of user $u$ and item $i$, respectively. 

Despite these advances, existing approaches largely fall into two categories: those that rely on semantic embeddings while overlooking the embedding size mismatch and limited task relevance, and those that focus on direct interaction prediction, which is computationally expensive. The former sacrifices structural integration opportunities, while the latter often struggles in complex real-world scenarios due to the challenges discussed above.  
\vspace{-2mm}

\section{Methodology}
\label{headings}
%In this section, we detail our proposed AGCF framework. As illustrated in Figure~\ref{fig:pipeline}, AGCF comprises two primary components: the \textbf{U-I-A Graph Construction} module and the \textbf{Behavioral Contrastive Learning} module.
In this section, we present the motivation and proposed \textbf{T}opology-\textbf{A}ugmented \textbf{G}raph \textbf{C}ollaborative \textbf{F}iltering (TAGCF) framework in detail. TAGCF consists of two key components: the \textbf{U-A-I Graph Construction} module and the \textbf{Adaptive Relation-weighted Graph Convolution} module.

\subsection{U-A-I Graph Construction}
To unlock the full potential of graph topology, we evolve the conventional User-Item bipartite graph into a tripartite User-Attribute-Item (U-A-I) structure. By introducing LLM-derived \textit{Reason} nodes as bridges, we create new indirect paths between users and items, achieving enhanced connectivity without the overhead of heavy text encoders during the inference stage.
In the U-A-I graph, auxiliary information is introduced through new paths via intermediary nodes, which gives semantically similar nodes an entirely new message-passing mechanism on the graph.
As illustrated in Figure \ref{fig:compare}c, user $u_1$ and item $i_2$ both share attribute $a_1$, indicating potential shared information. This results in a new path $u_1 \rightarrow a_1 \rightarrow i_2$, which allows $u_1$ to aggregate information from the embedding of $i_2$, and vice versa. This facilitates richer representation learning by enabling indirect information flow through shared attributes. 

Directly generating or removing user-item interactions often introduces significant noise due to the difficulty LLMs face when retrieving specific items from a vast candidate space. However, by introducing intermediary nodes, our approach mitigates this issue, shifting the model's task from high-cardinality item prediction to more manageable reasoning-based connectivity. Furthermore, the original user-item interaction information is fully preserved in the U-A-I graph, which ensures that the contribution of the original interaction data to the final node representations is fully retained.

The U-A-I graph construction is built upon two key modules. The first is the Attributes Extraction module, responsible for extracting raw attributes, while the second is the Filtering and Fusion module, which handles the filtering and fusion of the raw attributes. In the following, we will detail these two modules sequentially.

\vspace{-3pt}
\subsubsection{Attributes Extraction}

%We now detail the Attribute Extraction Module, which utilizes LLMs to infer latent causal attributes underlying user-item interactions. 

We now present the \textbf{Attribute Extraction} module, which leverages LLMs to infer latent causal attributes underlying user-item interactions. 

Given an interaction pair $\langle u,i \rangle$, user's review on the item as well as the items' metadata, the LLM leverages its extensive world knowledge to reason about the potential causes of the interaction and extract a set of semantically meaningful attributes ${\mathbf{a}_{u,i}}$ for both the user and the item. Specifically, we provide the LLM with the following prompts: (1) Set the system role and a scenario description tailored to the dataset context;
(2) A list of potential hidden attribute topics and curated domain-specific background knowledge;
(3) Several real-world examples annotated by human experts;
(4) The user-generated review $r$ related to item $v$ combined with item metadata.

To facilitate effective attribute extraction, we design tailored prompts for each dataset. These prompts provide topic guidance and illustrative examples, while still allowing for open-ended attribute generation without imposing rigid constraints. Since only a single interaction and its associated metadata are fed into the prompt at a time, the context length limitations of the LLMs are not a concern. Examples of the prompt can be found in our released code.

\begin{figure*}[h!]
    \centering
    \includegraphics[width=0.9\linewidth]{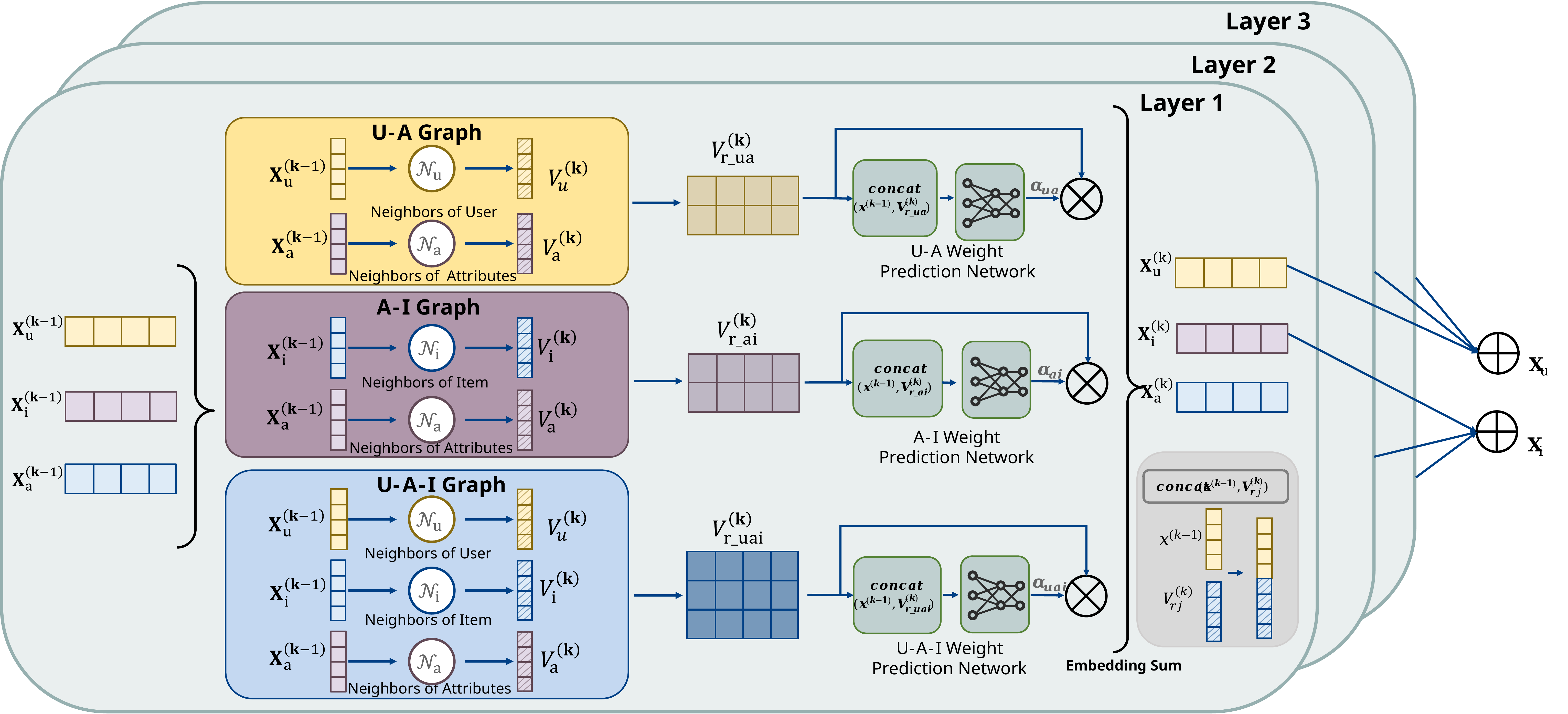}
    \caption{An illustration of the ARGC with adaptive weight prediction.}
    \label{fig:RGCN}
    \vspace{-15pt}
\end{figure*}

\subsubsection{Raw Attributes Filtering and Fusion}
To fully exploit the rich but noisy attribute knowledge extracted by LLMs, we design a Filtering and Fusion module that eliminates redundancy and consolidates semantically equivalent attributes before integrating them into the user–item interaction graph. This step is crucial because the raw attribute space is often excessively large, containing overlapping or weakly informative signals that may overwhelm the original collaborative filtering structure. By carefully filtering and fusing attributes, we effectively reduce sparsity and noise, preserve only the most representative signals, and alleviate the computational burden of subsequent learning. 

% %% 伪代码
% \begin{algorithm}[t]
% \caption{Filter and Fusion Module}
% \KwIn{Raw attribute set $\mathcal{A}_{raw}$, thresholds $\tau_{min}, \tau_{max}$, pretrained language model $LM$}
% \KwOut{Filtered and fused attribute set $\mathcal{A}$}

% \BlankLine
% \textbf{Step 1: Frequency-based Filtering} \\
% Initialize $\mathcal{A} \leftarrow \emptyset$ \;
% \ForEach{$a \in \mathcal{A}_{raw}$}{
%     \If{$\tau_{min} \leq \text{count}(a) \leq \tau_{max}$}{
%         $\mathcal{A} \leftarrow \mathcal{A} \cup \{a\}$ \;
%     }
% }

% \BlankLine
% \textbf{Step 2: Greedy Semantic Fusion} \\
% Sort attributes in $\mathcal{A}$ into a list $L$ \;
% Initialize $\mathcal{A}_{fused} \leftarrow \emptyset$ \;
% \While{$L$ not empty}{
%     Take first attribute $a_i$ from $L$ \;
%     Initialize cluster $C \leftarrow \{a_i\}$ \;
%     \ForEach{$a_j \in L$}{
%         \If{$LM.\text{MutualInclusion}(a_i, a_j)$ is True}{
%             $C \leftarrow C \cup \{a_j\}$ \;
%             Remove $a_j$ from $L$ \;
%         }
%     }
%     Choose $a_i$ as canonical form of $C$ \;
%     $\mathcal{A}_{fused} \leftarrow \mathcal{A}_{fused} \cup \{a_i\}$ \;
% }
% \Return{$\mathcal{A}_{fused}$} \;
% \end{algorithm}

\textbf{1. Frequency-based Filtering.}
We first apply a global frequency filter to prune attributes with extremely low or extremely high occurrence counts, as both ends are typically uninformative: rare attributes are too sparse to contribute meaningful patterns, while overly common attributes tend to act as constants. Formally, an attribute $a$ is retained only if its frequency lies within a predefined interval $[\tau_{\min}, \tau_{\max}]$:
\vspace{-5pt}
\begin{equation}
    \mathcal{A} = \{ a \mid  \tau_{min } \leq \text{count}(a) \leq \tau_{max}  \}.
\end{equation}

\vspace{-5pt}

\textbf{2. Greedy Semantic Fusion}. After frequency filtering, we further consolidate semantically redundant attributes. Directly clustering all attributes into semantic groups in a single step is infeasible, and exhaustively checking pairwise equivalence among all nodes would incur prohibitive computational costs.
To efficiently address this issue, we design a greedy fusion algorithm based on mutual semantic inclusion. Following prior work on lexical semantic equivalence, we employ a pretrained language model to determine whether two attribute concepts bidirectionally entail each other. Specifically, for each attribute $a_i$, we iteratively test mutual inclusion with subsequent attributes $a_j$. If equivalence is confirmed, $a_j$ is merged into the cluster of $a_i$, which is designated as the canonical representative. This process continues until no further merges are possible, yielding a compact set of unique and semantically non-overlapping attributes. The edges originally connected to redundant nodes are reassigned to their canonical representatives, with duplicate edges removed. Consequently, we obtain both a refined attribute vocabulary with independent semantics and a de-duplicated User-Attribute-Item Graph. The time saved by Greedy Semantic Fusion is shown on Table ~\ref{tab:datasets}.

\vspace{2pt}
\subsection{Adaptive Relation-weighted Graph Convolution}
Subsequent to attribute extraction and preprocessing, the critical challenge lies in effectively integrating the LLM-derived auxiliary information into the collaborative filtering signals. Prior approaches often adopt a direct fusion strategy, such as initializing embeddings with textual features or applying contrastive alignment. However, these methods treat the graph structure as static and homogeneous, overlooking the \textit{relational heterogeneity} within the User-Attribute-Item (U-A-I) graph. Consequently, they fail to exploit the structural priors where different types of connections (e.g., explicit user interactions vs. inferred attribute links) contribute differently to user preference modeling. To bridge this gap, we propose \emph{Adaptive Relation-weighted Graph Convolution} (ARGC), a mechanism explicitly designed to model relation-specific contributions and filter noise during message passing.

\subsubsection{Relation-Specific Aggregation}
To fully exploit attribute information while mitigating the impact of potential noise, ARGC convolves over three distinct relation-based graphs: $\mathbf{G}_{UAI}$ (User-Attribute-Item), $\mathbf{G}_{UA}$ (User-Attribute) and $\mathbf{G}_{IA}$ (Item-Attribute). 

Formally, for each relation type $r_j \in \mathcal{R} = \{UAI, UA, AI\}$, we decouple the message-passing process to maintain representation purity. Instead of a mixed aggregation, we first compute a \textit{relation-specific intermediate representation} $\mathbf{v}_{r_j}^{(k+1)}$ using the node embeddings $\mathbf{x}^{(k)}$ from the $k$-th layer. In each layer of these graphs, node embeddings are updated by aggregating messages from their respective neighborhoods:
\begin{equation}
\begin{split}
\mathbf{v}_{r_{j}}^{(k+1)} &=
\sum_{u \in \mathcal{N}_v^{(U)}}
\frac{1}{\sqrt{|\mathcal{N}_v|}\sqrt{|\mathcal{N}_u|}} \mathbf{x}_u^{(k)} 
+ \sum_{a \in \mathcal{N}_v^{(A)}}
\frac{1}{\sqrt{|\mathcal{N}_v|}\sqrt{|\mathcal{N}_a|}} \mathbf{x}_a^{(k)} \\
&+ \sum_{i \in \mathcal{N}_v^{(I)}}
\frac{1}{\sqrt{|\mathcal{N}_v|}\sqrt{|\mathcal{N}_i|}} \mathbf{x}_i^{(k)},
\end{split}
\label{eq:update_general}
\vspace{-5mm}
\end{equation}
\noindent where $\mathcal{N}_v^{(U)}$, $\mathcal{N}_v^{(I)}$, and $\mathcal{N}_v^{(A)}$ represent the subsets of user, item, and attribute neighbors directly connected to node $v$, respectively. Here, $\mathcal{N}_v = \mathcal{N}_v^{(U)} \cup \mathcal{N}_v^{(I)} \cup \mathcal{N}_v^{(A)}$ denotes the full graph node set. Take an attribute node in the U-A graph as an example. As it does not connect to other attribute or item nodes, the representation after aggregation is $\sum_{u \in \mathcal{N}_a}
\frac{1}{\sqrt{|\mathcal{N}_a|}\sqrt{|\mathcal{N}_u|}} \mathbf{x}_u^{(k)} $ .

\subsubsection{Adaptive Weighting and Noise Filtration}
To dynamically balance the contribution of each relation and mitigate the impact of potentially noisy attributes extracted by LLMs, we employ a lightweight prediction network. This network acts as a \textit{semantic gate}, estimating the importance of each relation-specific update based on its consistency with the node's current state. The adaptive weight is computed as:

{\footnotesize
\begin{equation}
\begin{split}
\alpha_{v, r_j} &= f(\mathbf{x}^{(k)}, \mathbf{v}_{r_j}^{(k+1)} \mid \theta_j) \\
&= W_{j_{2}}\!\left(\mathrm{LeakyReLU}\!\left(W_{j_{1}} \!\cdot\! 
\mathrm{concat}(\mathbf{x}^{(k)}, \mathbf{v}_{r_j}^{(k+1)}) 
+ b_{j_{1}}\right)\right) + b_{j_2},
\end{split}
\label{eq:relation_weight}
\end{equation}
}
\noindent where $\alpha_{v, r_j}$ serves as a scalar importance score for relation $r_j$ regarding node $v$. 
Crucially, the prediction network $f(\cdot)$ conditions on both the current node representation $\mathbf{x}^{(k)}$ and the candidate update $\mathbf{v}_{r_j}^{(k+1)}$. This design allows the model to assess the \textbf{information gain} and \textbf{semantic consistency} provided by the relation. For example, if an LLM-inferred attribute link (in $\mathbf{v}_{UA}^{(k+1)}$) contradicts the user's established preference history ($\mathbf{x}^{(k)}$), the network can assign a lower weight to prune this noisy reasoning path.

The final node representation for layer $k+1$ is obtained by fusing these weighted relation-specific signals:
\begin{equation}
\mathbf{x}^{(k+1)} = \sum_{r_j \in \mathcal{R}} 
\alpha_{v, r_j} \cdot \mathbf{v}_{r_j}^{(k+1)}.
\end{equation}

Finally, we perform layer combination to obtain the final embedding $\mathbf{x} = \sum_{k=0}^K \mathbf{x}^{(k)}$. The model is optimized using the Bayesian Personalized Ranking (BPR) loss augmented with regularization terms:
\begin{equation}
\mathcal{L} = - \sum_{(u,i,j) \in \mathcal{D}} 
\log \sigma \big( \hat{y}_{ui} - \hat{y}_{uj} \big)
+ \lambda \| \Theta \|_2^2  , 
\label{eq:bpr_loss}
\end{equation}
\noindent where $\lambda$ control the strength of $L_2$ regularization.
% and possible different kind of contrastive loss, respectively.

% \subsection{Discussion: Distinction from Existing Methods}
\subsection{Comparative Analysis}
In this section, we delineate the methodological distinctiveness of our approach by contrasting it with established baselines and discussing how it overcomes their inherent limitations. To ground the aforementioned discussion in a formal framework, consider the standard representation where the set of users and items are denoted by $\mathcal{U}=\{u_1,...u_I\}$ and $\mathcal{I}=\{i_1,...i_J\}$, respectively. A user-item interaction is denoted by the pair $\langle u, i \rangle$, indicating that user $u$ has interacted with item $i$. 
% We use $I_x$ to represent the set of items interacted with by user $u_x$.
%We denote by $I_x$ the set of items interacted by user $u_X$.
Before training, users and items are initialized with random embeddings $\mathbf{X}_U = \{\mathbf{x}_{u1},..., \mathbf{x}_{uN}\}$ and $\mathbf{X}_I = \{\mathbf{x}_{i1},...\mathbf{x}_{iJ}\}$, 
respectively, and are subsequently aggregated via an adjacency matrix derived from the User-Item Interaction Graph $\mathbf{G}_{UI}$. The aggregated process is:
\begin{equation}
    \mathbf{G}_{UI} = \begin{bmatrix}
        0 &\mathbf{G}_{UI} \\
        \mathbf{{G}}_{UI}^{T} & 0
    \end{bmatrix},
\end{equation}
\begin{equation}
    \mathbf{X}^{t} = (D^{-\frac{1}{2}} \mathbf{G}_{UI} D^{\frac{1}{2}}) \mathbf{X}^{t-1} ,
    \label{eq:aggregate_ui}
\end{equation}

\noindent where the $ D^{-\frac{1}{2}} \mathbf{G}_{UI} D^{\frac{1}{2}}$ represents the normalized adjacency matrix of the graph $\mathbf{G}_{UI}$.

\textbf{Contrast with Existing LLM-based Augmentation.}
Prior efforts to integrate LLMs into CF typically optimize along two orthogonal axes: \textit{Structure Refinement} and \textit{Feature Enrichment}.

\textit{(i) Structure Refinement (Edge Augmentation):} These methods aim to mitigate sparsity by using LLMs to predict missing interactions, effectively modifying the adjacency matrix $\mathbf{G}_{UI}$ directly:
\vspace{-2pt}
\begin{equation}
    \mathbf{G}'_{UI} = \mathbf{G}_{UI} + \mathbf{G}_{LLM},
\end{equation}
\vspace{-2pt}
where $\mathbf{G}_{LLM}$ represents pseudo-edges generated by the LLM. However, directly predicting specific items from a massive candidate pool is prone to hallucinations, introducing significant noise into the graph structure.

\textit{(ii) Feature Enrichment (Embedding Augmentation):} These methods enhance representations by injecting external textual embeddings $\mathbf{X}_{text}$ (LLM embeddings) into the user/item features:
\begin{equation}
    \mathbf{X} = \mathrm{Fusion}(\mathbf{X}_{id}, \mathbf{X}_{text}).
\end{equation}
While semantically rich, this approach suffers from \textit{dimension mismatch}, forcing high-dimensional linguistic features into a low-dimensional collaborative space, and treats LLM knowledge as static attributes rather than structural signals.

In contrast to these paradigms, TAGCF argues that semantic knowledge should be represented as \textit{connectivity} rather than features. We neither blindly add noisy $\langle u, i \rangle$ edges nor rely on heavy text features. Instead, we introduce intermediate attribute nodes $\mathcal{A}$ to explicitly bridge users and items.
This fundamentally transforms the graph topology from a bipartite structure to a tripartite one, expanding the adjacency matrix as follows:
\begin{equation}
\mathbf{G}_{UAI} = \begin{bmatrix}
\mathbf{0} & \mathbf{G}_{UI} & \mathbf{G}_{UA} \\
\mathbf{G}_{UI}^\top & \mathbf{0} & \mathbf{G}_{IA} \\
\mathbf{G}_{UA}^\top & \mathbf{G}_{IA}^\top & \mathbf{0}
\end{bmatrix}.
\end{equation}
Consequently, the aggregation process is characterized by:
\begin{equation}
    \mathbf{X}^{(l+1)} = \sigma\left( \tilde{\mathbf{G}}_{UAI} \mathbf{X}^{(l)} \right).
    \label{eq:aggregate_uai}
\end{equation}
This formulation highlights our core distinction: TAGCF transforms ``semantic reasoning'' into ``topological path'' (e.g., $u \to a \to i$). By initializing attribute nodes with learnable IDs, we capture interaction causes through structural message passing alone, avoiding the intractability of direct edge prediction and the dimensional misalignment of text embeddings.

\textbf{Contrast with Knowledge Graph (KG)-based Methods.}
While KG-based methods also utilize auxiliary graphs (typically Item-Entity relations), ARGC distinguishes itself in three critical aspects:
\textit{Unified vs. Disjoint Integration}: KG methods often treat the collaborative graph ($\mathbf{G}_{UI}$) and knowledge graph ($\mathbf{G}_{KG}$) as separate views, merging them loosely via multi-task learning. In contrast, ARGC integrates all relations into a unified $\mathbf{G}_{UAI}$ topology, allowing for seamless information flow across different relation types.
\textit{User-Centric Bridging}: Conventional KGs typically connect items to external entities (Item-Entity-Item paths). We introduce a novel User-Attribute subgraph, creating explicit User-Attribute-Item paths that interpret {why} a user interacts with an item.
\textit{Relevance vs. Noise}: External KGs often contain generic, task-irrelevant relations. Our attributes are extracted specifically for the recommendation context, ensuring they are discriminative for preference modeling.

\textbf{Contrast with Previous Graph Mechanisms.} It is important to note that while ARGC involves adaptive weighting, it differs fundamentally from standard graph attention mechanisms (e.g., GAT\cite{GAT}). Standard attention computes pair-wise coefficients for every edge, which incurs high computational costs of $O(|E|)$ and is prone to overfitting on sparse, noisy interaction graphs. In contrast, multi-relational frameworks like RGCN\cite{rgcn} apply a globally shared transformation to all edges within the same relation type. Consequently, they lack the flexibility to capture fine-grained, instance-specific variations underlying each interaction.
In contrast, ARGC operates at the \textit{relation level}, learning to gate broad categories of information (e.g., ``how much should I trust attribute reasoning vs. historical behavior for this user?''). This structural constraint acts as a regularization prior, making our approach more robust and efficient for collaborative filtering tasks.

In summary, by unifying multi-relational structures, introducing user–attribute bridges, and relation-level adaptive weighting, TAGCF fundamentally differs from traditional knowledge graph–based approaches, achieving more robust representation learning and better alignment with collaborative signals.

\vspace{-1mm}

\section{Evaluation}
In this section, we conduct comprehensive experiments to validate the effectiveness of our proposed TAGCF framework when integrated with several widely-used backbone models. Subsequently, experiments demonstrating how our method addresses data sparsity and some case studies are provided. In addition, we present further analyses, including ablation studies and deep analysis of the U-A-I graph. Moreover, we provide a detailed case study of Amazon-Office dataset.

\label{datasets}

\begin{table}[t!]
\centering
\vspace{-5pt}
\caption{Dataset information.}
\vspace{-5pt}
\resizebox{\linewidth}{!}{
\begin{tabular}{c|ccc}
\hline
\textbf{Statistic} & \textbf{Books} & \textbf{Yelp} & \textbf{Office} \\ \hline
Users              & 11,000         & 11,091        & 6,338           \\ 
Items              & 9,332          & 11,010        & 4,461           \\ 
Attributes         & 1,978          & 2,607         & 4,065           \\ 
U-I Interactions   & 120,464        & 166,620       & 56,190          \\ 
Density            & 1.2e-3         & 1.4e-3        & 2.0e-3          \\ \hline
API Costs (\$)      & 76.1           & 90.7          & 38.8            \\
FF Time w/o Greedy Search (h) & 87 & 110 & 31  \\
FF Time (h) & 5.9 & 6.5 & 2.8 \\ \hline

\end{tabular}
}
\vspace{-10pt}
\label{tab:datasets}
\end{table}

\subsection{Experimental Settings}

\begin{table*}[h]
\centering
\caption{Recommendation performance Improvement of all backbone methods on different datasets in terms of Recall and NDCG. Values in bold represent the best performance.}
\label{tab:main_table}
\setlength{\tabcolsep}{2.5pt} 
% \resizebox{\textwidth}{!}{
\begin{tabular}{ll *{12}{S[table-format=1.4]}} % 使用 S 列类型对齐小数点
\toprule
\multirow{2}{*}{\textbf{Dataset}} &  & \multicolumn{4}{c}{\textbf{Amazon-Book}} & \multicolumn{4}{c}{\textbf{Yelp}} & \multicolumn{4}{c}{\textbf{Amazon-Office}} \\
\cmidrule(lr){3-6} \cmidrule(lr){7-10} \cmidrule(lr){11-14} % 使用 cmidrule
 & & \multicolumn{2}{c}{Recall} & \multicolumn{2}{c}{NDCG} & \multicolumn{2}{c}{Recall} & \multicolumn{2}{c}{NDCG} & \multicolumn{2}{c}{Recall} & \multicolumn{2}{c}{NDCG} \\
\cmidrule(lr){1-2}\cmidrule(lr){3-4} \cmidrule(lr){5-6} \cmidrule(lr){7-8} \cmidrule(lr){9-10} \cmidrule(lr){11-12} \cmidrule(lr){13-14}
\textbf{CF-Model} & \textbf{Aug Method} & {@5} & {@20} & {@5} & {@20} & {@5} & {@20} & {@5} & {@20} & {@5} & {@20} & {@5} & {@20} \\
\midrule
\textbf{AutoCF} & - & {0.0688} & {0.1559} & {0.0714} & {0.0998} & {0.0476} & {0.1306} & {0.0548} & {0.0821} & {0.0299} & {0.0754} & {0.0288} & {0.0470} \\
\textbf{AlphaRec} & - & {0.0660} & {0.1535} & {0.0671} & {0.0978} & {0.0460} & {0.1289} & {0.0522} & {0.0780} & {0.0341} & {0.0897} & {0.0332} & {0.0551} \\
\midrule
\multirow{6}{*}{\textbf{LightGCN}} & - & 0.0626 & 0.1482 & 0.0630 & 0.0919 & 0.0436 & 0.1200 & 0.0507 & 0.0760 & 0.0336 & 0.0889 & 0.0320 & {0.0538} \\
 & $+\textit{KAR}$ & {0.0642} & {0.1491} & {0.0654} & {0.0942} & {0.0443} & {0.1223} & {0.0519} & {0.0774} & {0.0351} & {0.0900} & {0.0335} & {0.0554} \\
 & $+\textit{RLMRec-Con}$ & {0.0666} & {0.1561} & {0.0665} & {0.0965} & {0.0453} & {0.1267} & {0.0532} & {0.0798} & {0.0312} & {0.0801} & {0.0301} & { 0.0494} \\
 & $+\textit{RLMRec-Gen}$ & {0.0712} & {0.1615} & {0.0716} & {0.1013}  & {0.0416} & {0.1175} & {0.0490} & {0.0743} & {0.0354} & {0.0874} & { 0.0341} & {0.0547} \\
 & $+\textit{TAGCF}$ & {\textbf{0.0743}} & {\textbf{0.1670}} & {\textbf{0.0737}} & {\textbf{0.1047}} & {\textbf{0.0463}} & {\textbf{0.1299}} & {\textbf{0.0539}} & {\textbf{0.0817}} & {\textbf{0.0373}} & {\textbf{0.0978}} & {\textbf{ 0.0364}} & {\textbf{0.0601}} \\
 & Improv.(\%) & {18.69} & {12.69} & {21.90} & {13.93} & {8.25} & {7.92} & {6.31} & {7.50} & {11.01} & {8.79} & {13.75} & {11.71} \\
\midrule
\multirow{6}{*}{\textbf{SGL}} & - & {0.0736} & {0.1653} & {0.0730} & {0.1038} & {0.0446} & {0.1241} & {0.0518} & {0.0787} & {0.0362} & {0.0896} & {0.0354} & { 0.0564} \\
 & $+\textit{KAR}$ & {0.0744} & {0.1665} & {0.0739} & {0.1047} & {0.0457} & {0.1255} & {0.0527} & {0.0794} & {0.0350} & {0.0860} &{0.0355} & {0.0557} \\
 & $+\textit{RLMRec-Con}$ & {0.0754} & {0.1700} & {0.0739} & {0.1055} & {0.0475} & {0.1307} & {0.0554} & {0.0827} & {0.0347} & {0.0841} & {0.0346} & {0.0540} \\
 & $+\textit{RLMRec-Gen}$ & {0.0738} & {0.1645} & {0.0734} & {0.1038} & {0.0479} &{0.1288} & {0.0553} & {0.0819} & {0.0375} & {0.0891} & {0.0371} & {0.0578} \\
 & $+\textit{TAGCF}$ & {\textbf{0.0863}} & {\textbf{0.1843}} & {\textbf{0.0860}} & {\textbf{0.1186}} & {\textbf{0.0506}} & {\textbf{0.1368}} & {\textbf{0.0585}} & {\textbf{0.0871}} & {\textbf{0.0380}} & {\textbf{0.0993}} & {\textbf{0.0369}} & {\textbf{ 0.0609 }} \\
 & Improv.(\%) & {17.26} & {11.49} & {17.81} & {14.26} & {13.45} & {10.23} & {12.93} & {10.67} & {4.70} & {10.83} & {7.63} & {12.08} \\
\midrule
\multirow{6}{*}{\textbf{SimGCL}} & - & {0.0676} & {0.1627} & {0.0676} & {0.0944} & {0.0477} & {0.1287} & {0.0552} & {0.0818} & {0.0354} & {0.0873 } & {0.0347} & {0.0552} \\
 & $+\textit{KAR}$ & {0.0762} & {0.1673} & {0.0760} & {0.1068} & {0.0482} & {0.1285} & {0.0557} & {0.0815} & {0.0343} & {0.0857} & {0.0341} & {0.0536} \\
 & $+\textit{RLMRec-Con}$ & {0.0699} & {0.1657} & {0.0699} & {0.1018} & {0.0486} & {0.1329} & {0.0566} & {0.0842} & {0.0333} & {0.0814} & {0.0328} & {0.0519} \\
 & $+ \textit{RLMRec-Gen}$ & {0.0654} & {0.1575} & {0.0661} & {0.0970} & {0.0467} & {0.1303} & {0.0553} & { 0.0826} & {0.0349} & {0.0868} & {0.0337} & {0.0543} \\
 & $+\textit{TAGCF}$ & {\textbf{0.0820}} & {\textbf{0.1819}} & {\textbf{0.0819} } & {\textbf{0.1153}} & \textbf{0.0525} & \textbf{0.1409} & \textbf{0.0600} & \textbf{0.0890} & {\textbf{0.0365}} & {\textbf{0.0976}} & {\textbf{0.0361}} & {\textbf{0.0600}} \\
 & Improv.(\%) & {21.30} & {11.80} & {21.15} & {22.14} & {10.06} & {9.48} & {8.70} & {8.80} & {3.11} & {11.80} & {4.03} & {8.70} \\
\bottomrule
\end{tabular}
% }
\end{table*}
\textbf{Dataset}. Our experiments are conducted on three public benchmark datasets: Amazon Book, Amazon Office, and Yelp. The selection of these datasets is motivated by our model's reliance on user-generated reviews and the need for data diversity. Specifically, the Amazon datasets provide user-item interactions related to books and office supplies, while the Yelp dataset offers reviews for a wide range of local businesses. To ensure the quality of the datasets and a minimum level of user engagement, we adopt the preprocessing procedure from \cite{Ren_2024RLM, Wang2019NeuralGraphCollaborative}, applying a k-core filtering where $k=10$. Following AlphaRec~\cite{Ren_2024RLM}, we partition each dataset into training, validation, and testing sets using a 3:1:1 ratio. All attributes are extracted from the training set. Detailed statistics for each dataset are provided in Table~\ref{tab:datasets}. The FF in the table represents Filter and Fusion Model, and the time is estimated on 4x NVIDIA RTX 4090 GPUs. Greedy Semantic Fusion drastically cuts down the computational overhead, leading to a much shorter processing time.

\textbf{Backbones and Baselines}. To validate the versatility of our framework, we integrate it with three different GNN-based backbones: LightGCN~\cite{He2020LightGCN}, SGL\cite{wu2021self}, and SimGCL~\cite{10.1145/3477495.3531937SimGCL}. Our comparative analysis includes two LLM-enhanced baselines, KAR\cite{xi2024towards} and RLMRec\cite{Ren_2024RLM}, as well as a strong cf model AutoCF\cite{xia2023automated} . Furthermore, we also compare our model against another LLM-based method AlphaRec as an individual baseline. The embedding sizes of all models are set to $64$, and the patience for early stop during training is set to $5$. The super parameters of baselines follow the original paper. All attributes are generated using DeepSeek V3.1, and all models converge within less than one hour on a single NVIDIA RTX 4090 GPU. We employed DeBERTa-v2 with threshold 0.5 for the Greedy Semantic Fusion q. Since the attributes are generated prior to training, the overall time complexity of each model is determined solely by its inherent computational complexity. 

\textbf{Evaluation Metrics}. We evaluate the model's performance using two widely adopted ranking-based metrics: Recall@N and NDCG@N. These metrics measure the quality of the top-N recommendation list generated by the model. In the experiments, we set N to 5 and 20.

\vspace{-5pt}
\subsection{Performance Comparison}

% Table~\ref{tab:main_table} presents a comprehensive summary of AGCF's performance, with "Improv." quantifying the performance gains over the backbone methods. The empirical results unequivocally demonstrate that AGCF establishes new state-of-the-art (SOTA) benchmarks across all datasets and foundational CF models. The approach exhibits strong model-agnostic generalization, yielding good performance regardless of the backbone model.
% It is particularly noteworthy that AGCF, despite not utilizing explicit textual embeddings, consistently surpasses methods that depend on them. This achievement highlights the efficacy of our approach in capturing latent semantics of user preferences and item characteristics. Moreover, it validates the architectural elegance and effectiveness of seamlessly injecting this semantic knowledge directly into the user-item graph.
Table~\ref{tab:main_table} presents a comprehensive evaluation of TAGCF's performance across three benchmark datasets and multiple backbone models, with ''Improv.'' quantifying the relative performance gains over the respective backbone methods. The empirical results demonstrate that TAGCF establishes new state-of-the-art benchmarks across all experimental configurations, achieving the best performance on every dataset-backbone combination with substantial and consistent improvements. 
On Amazon-Book, TAGCF achieves improvements ranging from 11.49\% to 22.14\% across different metrics and backbones; on Yelp, the improvements span from 6.31\% to 13.45\%; and on Amazon-Office, the gains range from 3.11\% to 13.75\%.
These consistent improvements across diverse datasets with varying characteristics, including different sparsity levels and domain specificity, demonstrate the robustness and broad applicability of our approach. The performance gains exhibit interesting patterns across different domains and evaluation metrics. 

A particularly compelling aspect of our results is TAGCF's strong model-agnostic generalization capability. The framework consistently enhances performance regardless of the underlying backbone architecture, whether it's the foundational LightGCN, the self-supervised SGL, or the contrastive SimGCL. This universality suggests that our attribute-augmented graph structure and adaptive relation weighting mechanism address fundamental limitations in collaborative filtering rather than being tailored to specific model architectures. Notably, the improvements are most pronounced with LightGCN, with up to 21.90\% improvement in NDCG@5, indicating that our semantic enhancement is particularly valuable for simpler backbone models that may lack sophisticated feature learning capabilities.

Despite not utilizing explicit textual embeddings or direct LLM predictions, TAGCF consistently outperforms existing LLM-enhanced methods including KAR, both variants of RLMRec (Contrastive and Generative) and AlphaRec. This superiority is particularly significant as it demonstrates that our structural enhancement approach without semantic embedding is more effective than direct textual feature augmentation or LLM-based interaction synthesis. 
% For instance, on Amazon-Book with LightGCN, AGCF achieves 0.0743 Recall@5 compared to RLMRec-Gen's 0.0712, representing a meaningful improvement despite RLMRec's direct utilization of LLM-generated content. Similarly, on Yelp with SGL, AGCF outperforms RLMRec-Con by achieving 0.0585 NDCG@5 versus 0.0554, demonstrating the effectiveness of our semantic attribute injection approach.

\vspace{-2mm}
\subsection{Ablation Study}

\begin{table}[t]
\caption{Ablation study on Amazon-Book}
\vspace{-3mm}
\resizebox{1.0\linewidth}{!}{
\centering
\begin{tabular}{c|l|c|c|c|c}
\hline
\textbf{} & & \textbf{Recall@5} & \textbf{Recall@20} & \textbf{NDCG@5} & \textbf{NDCG@20} \\ \hline
\multirow{5}{*}{\textbf{Lightgcn}} & TAGCF & 0.0743 & 0.1670 & 0.0737 & 0.1047 \\ 
& w/o ARGC & 0.0688 & 0.1569 & 0.0690 & 0.0985 \\ 
& w/o FF & 0.0628 & 0.1461 & 0.0605 & 0.0890 \\ 
& w/o FF \& ARGC & 0.0661 & 0.1537 & 0.0656 & 0.0948 \\ 
& Base & 0.0626 & 0.1482 & 0.0630 & 0.0919 \\ \hline
\multirow{5}{*}{\textbf{SGL}} & TAGCF & 0.0863 & 0.1843 & 0.0860 & 0.1186 \\ 
& w/o ARGC & 0.0771 & 0.1676 & 0.0756 & 0.1058 \\ 
& w/o FF & 0.0686 & 0.1540 & 0.0693 & 0.0975 \\ 
& w/o FF \& ARGC & 0.0735 & 0.1655 & 0.0726 & 0.1038 \\ 
& Base & 0.0736 & 0.1653 & 0.0730 & 0.1038 \\ \hline
\multirow{5}{*}{\textbf{SimGCL}} & TAGCF & 0.082 & 0.1819 & 0.0819 & 0.1153 \\ 
& w/o ARGC & 0.0783 & 0.1740 & 0.0780 & 0.1103 \\ 
& w/o FF & 0.0628 & 0.1461 & 0.0605 & 0.0890 \\ 
& w/o FF \& ARGC & 0.0741 & 0.1537 & 0.0656 & 0.0944 \\ 
& Base & 0.0676 & 0.1627 & 0.0676 & 0.0944 \\ \hline
\end{tabular}

}

\vspace{-10pt}
\label{tab:ablation}
\end{table}

To investigate the contribution of each key component in TAGCF, we perform an ablation study on the Amazon datasets with two base models, Lightgcn and SimGCL. Specifically, we gradually remove main modules: the Adaptive Relation-weighted Graph Convolution (ARGC), the Filter and Fusion (FF) module and both. The results are reported in Table~\ref{tab:ablation}.

From the results, we can draw several observations.  
First, removing the ARGC module consistently degrades performance across all models, with particularly notable drops in NDCG scores. For LightGCN, removing ARGC causes a 6.38\% decrease in NDCG@5 and 5.92\% in NDCG@20. This pattern indicates that adaptive relation weighting primarily enhances ranking quality, addressing fundamental limitations in handling heterogeneous edge types.
Second, the Filter and Fusion module removal produces the biggest performance drops. With SGL, removing FF causes a 20.47\% decrease in NDCG@5 and 17.79\% in NDCG@20, representing the largest single-component impact. This substantial degradation underscores that raw LLM-extracted attributes require sophisticated integration mechanisms to align effectively with collaborative filtering signals.
Interestingly, simultaneous removal of both components (w/o FF \& ARGC) sometimes outperforms single-component removal variants. With LightGCN, w/o FF \& ARGC achieves better performance than w/o FF alone, suggesting that ARGC may be less effective when attribute integration is fundamentally flawed.
Finally, complete TAGCF consistently achieves superior performance across all metrics, with improvements ranging from 5.7\% to 27.2\% over ablated variants. This validates that the synergistic combination of adaptive relation weighting and attribute integration is essential for maximizing LLM-enhanced collaborative filtering benefits.

For $\tau_{min}$ and $\tau_{max}$, we conduct several experiments, and we found that varying $\tau_{min}$ range [3, 10] and $\tau_{max}$ range [1000, 10000] has negligible impact. However, setting $\tau_{min} <3$ leads to a ~5\% to 10\% performance drop, validating our strategy.

% First, removing the ARGC module (\textit{w/o ARGC}) consistently leads to a noticeable drop in both \textit{Recall} and \textit{NDCG}, confirming that the adaptive relation weighting mechanism effectively enhances information propagation by distinguishing the importance of different edge types.  
% Second, eliminating the Filter and Fusion module (\textit{w/o FF}) causes the most significant degradation, which highlights the necessity of attributes preprocessing. 

% Overall, the complete AGCF achieves the best results across all metrics, validating that each designed component contributes synergistically to the model’s final performance.

\subsection{ Data Sparsity \& Cold Start }
A primary limitation of conventional CF recommenders is their vulnerability to data sparsity and the cold-start problem, as embeddings learned purely from interactions are inept at handling novel items. To rigorously assess TAGCF's resilience to these challenges, we conducted dedicated experiments on the Amazon-Books dataset, employing LightGCN as the backbone.

\begin{figure}
    \centering
    % 关键修改：直接在 figure* 中使用 subfloat 和 \textwidth 比例
    
    % *** 第一行（两张图）***
    % \resizebox{1.0\linewidth}{!}{
    
    \subfloat[Recall@20  vs Sparse Ratio.]{
        % 每张图占总宽度（\textwidth）的 48%
        \includegraphics[width=0.48\linewidth]{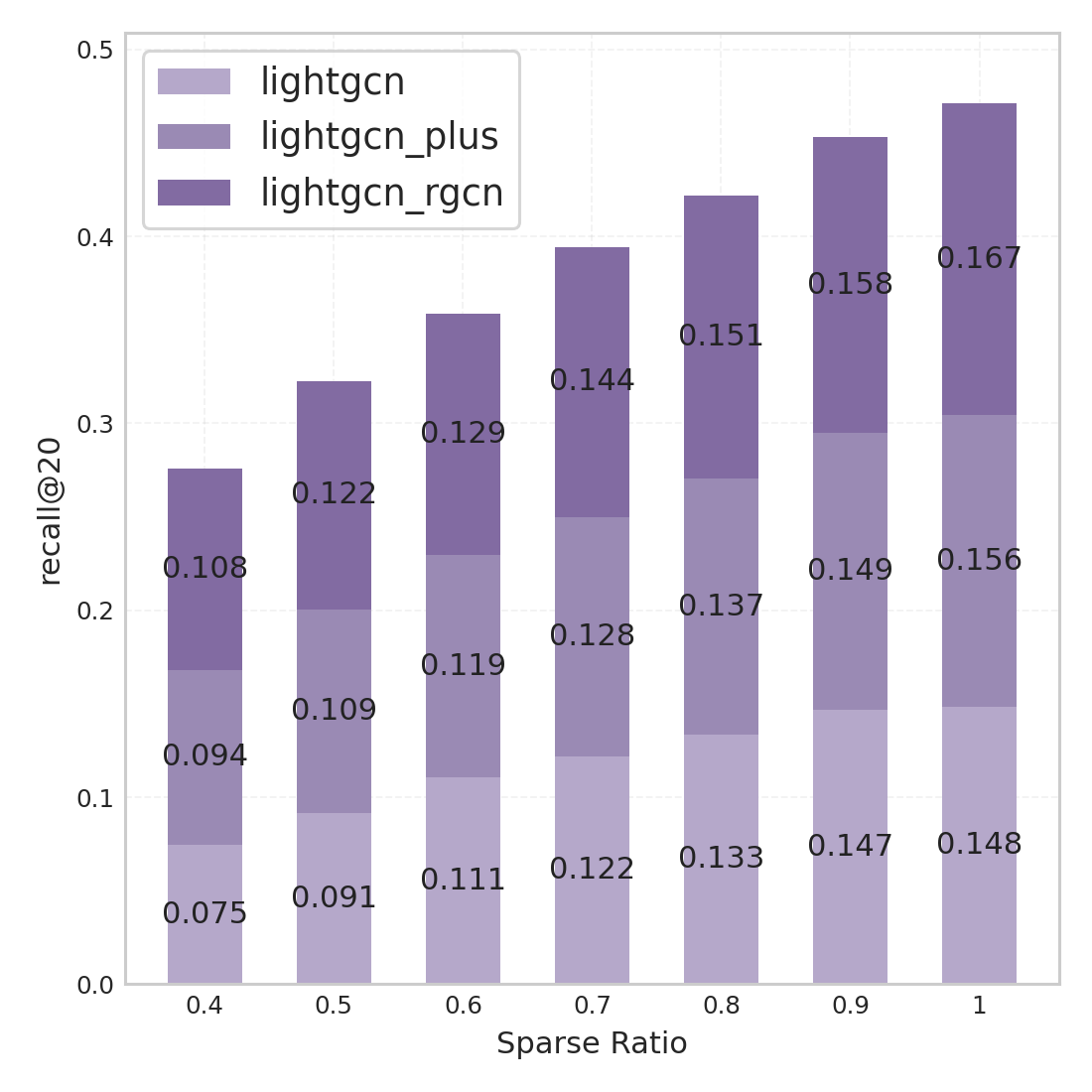} 
        \label{fig:sparse_fig1}
    }
    \hfill
    \subfloat[NDCG@20 vs Sparse Ratio.]{
        \includegraphics[width=0.48\linewidth]{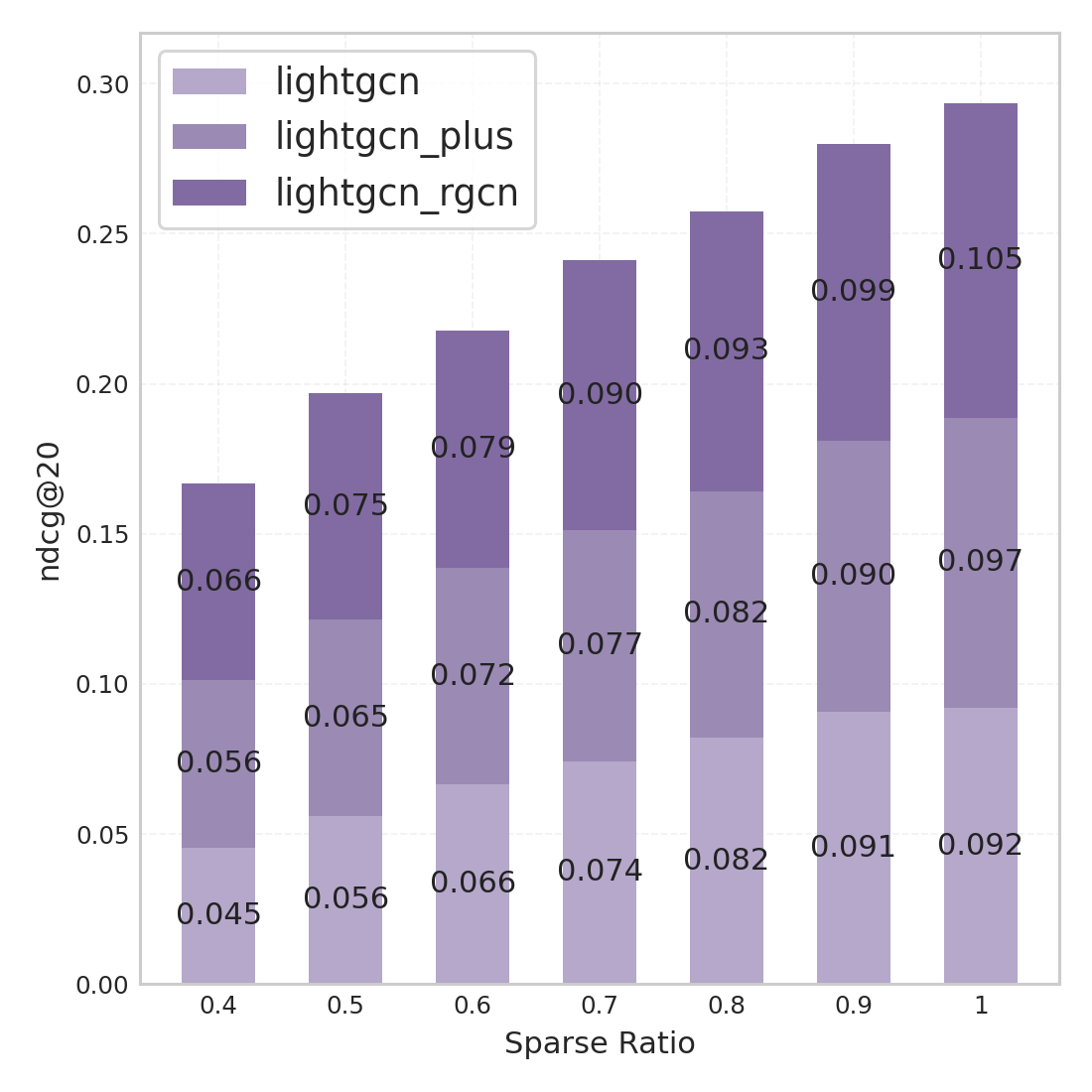}
        \label{fig:sparse_fig2}
    }
    
    \vspace{-6pt} % 垂直间距
    
    % *** 第二行（两张图）***
    \subfloat[Recall@20 vs Cold Start Ratio.]{
        \includegraphics[width=0.48\linewidth]{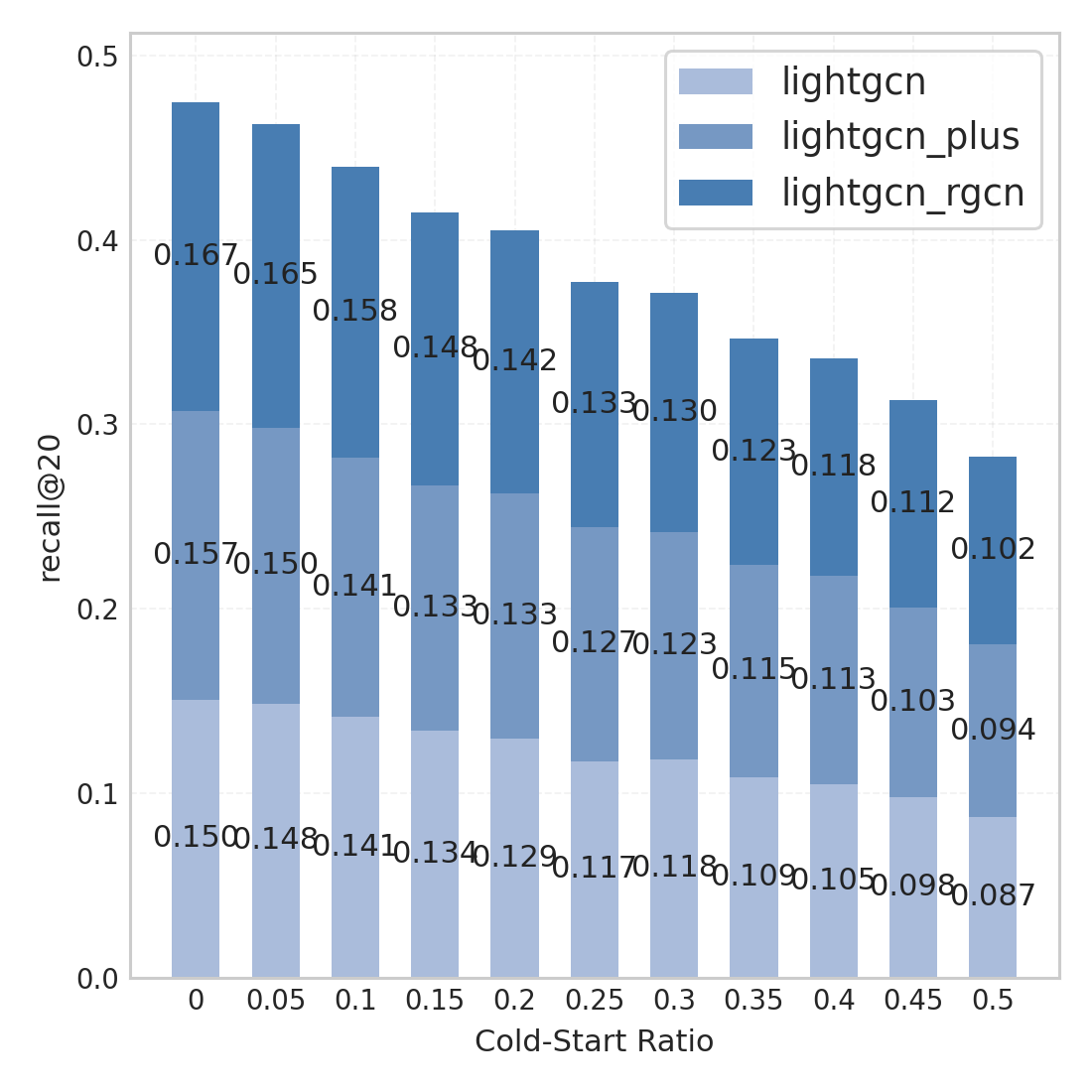}
        \label{fig:cold_start_fig1}
    }
    \hfill
    \subfloat[NDCG@20  vs Cold Start Ratio.]{
        \includegraphics[width=0.48\linewidth]{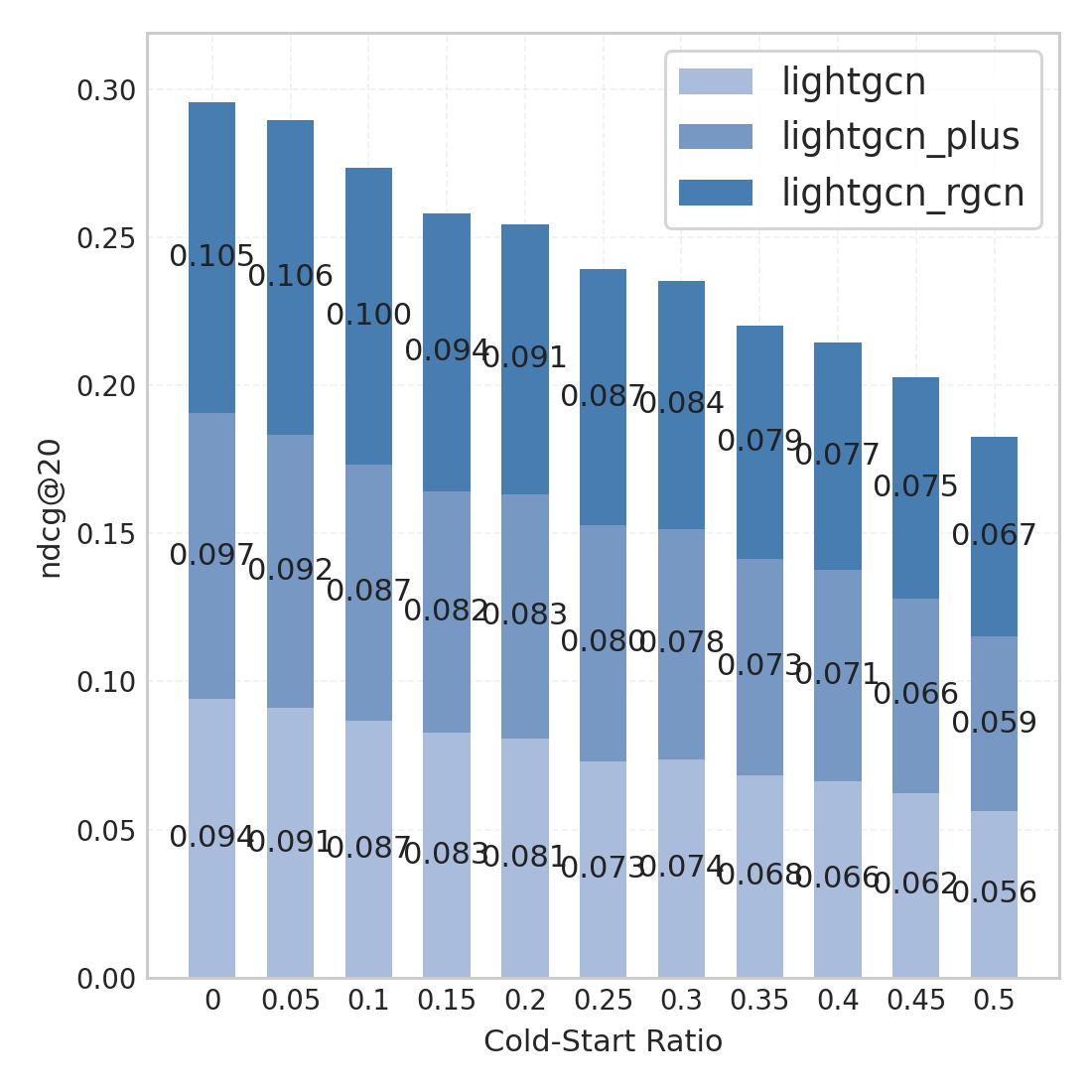}
        \label{fig:cold_start_fig2}
    }
% }
    \caption{Data sparsity and cold start experiments.}
    \label{fig:all_results}
    \vspace{-15pt}
\end{figure}

\textbf{Data Sparsity.}
To simulate varying degrees of data scarcity, we systematically pruned the training set by randomly removing user-item interactions based on a retention ratio, $s$, while the validation and test sets remained intact. The ratio $s$ is formally defined as:
$
    s = |<u,i>_{retained}|/|<u,i>_{original}|,
$  where a lower $s$ signifies a greater degree of sparsity and a more formidable recommendation task. The results, illustrated in Figures~\ref{fig:sparse_fig1} and \ref{fig:sparse_fig2}, reveal that TAGCF maintains a commanding performance advantage over both the vanilla LightGCN backbone and the text-enhanced RLMRec model (labeled as LightGCN-plus) across all evaluated sparsity levels. Impressively, TAGCF operating on just 70\% of the interaction data matches the performance of a fully-trained LightGCN. This provides compelling evidence that the auxiliary semantic information harnessed by TAGCF serves as a powerful mitigator for the data sparsity issue.

\begin{figure*}[htbp]
\vspace{-1mm}

    \centering
    \resizebox{1.0\linewidth}{!}{
    % 关键修改：直接在 figure* 中使用 subfloat 和 \textwidth 比例
    % *** 第一行（两张图）***
    \subfloat[\fontsize{12pt}{13pt}\selectfont Amazon-Book Recall@20 vs Layer Num.]{
        % 每张图占总宽度（\textwidth）的 48%
        \includegraphics[width=0.48\textwidth]{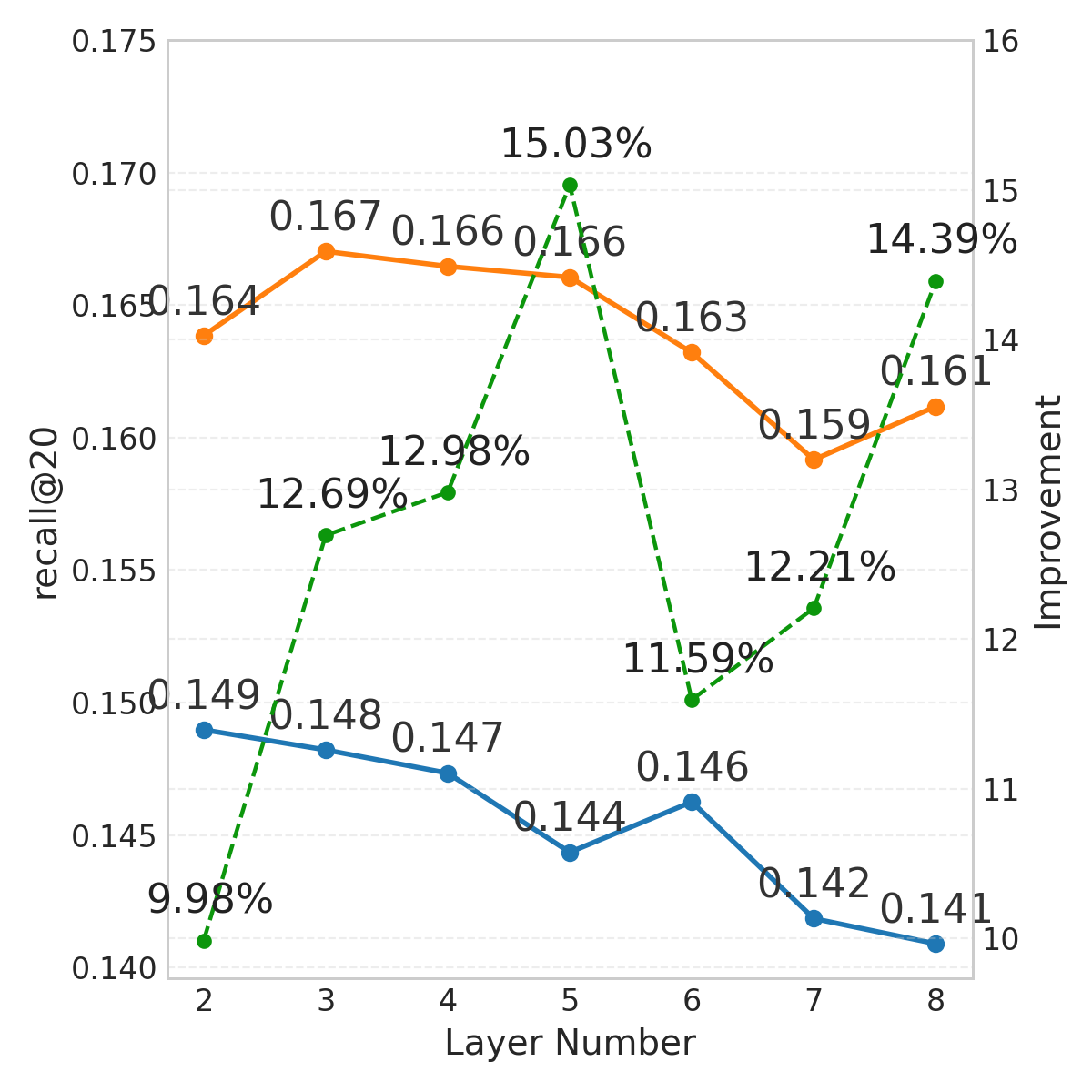} 
        \label{fig:amazon_recall_20_vs_layer}
    }
    \hfill
    \subfloat[\fontsize{12pt}{13pt}\selectfont Amazon-Book NDCG@20 vs Layer Num.]{
        \includegraphics[width=0.48\textwidth]{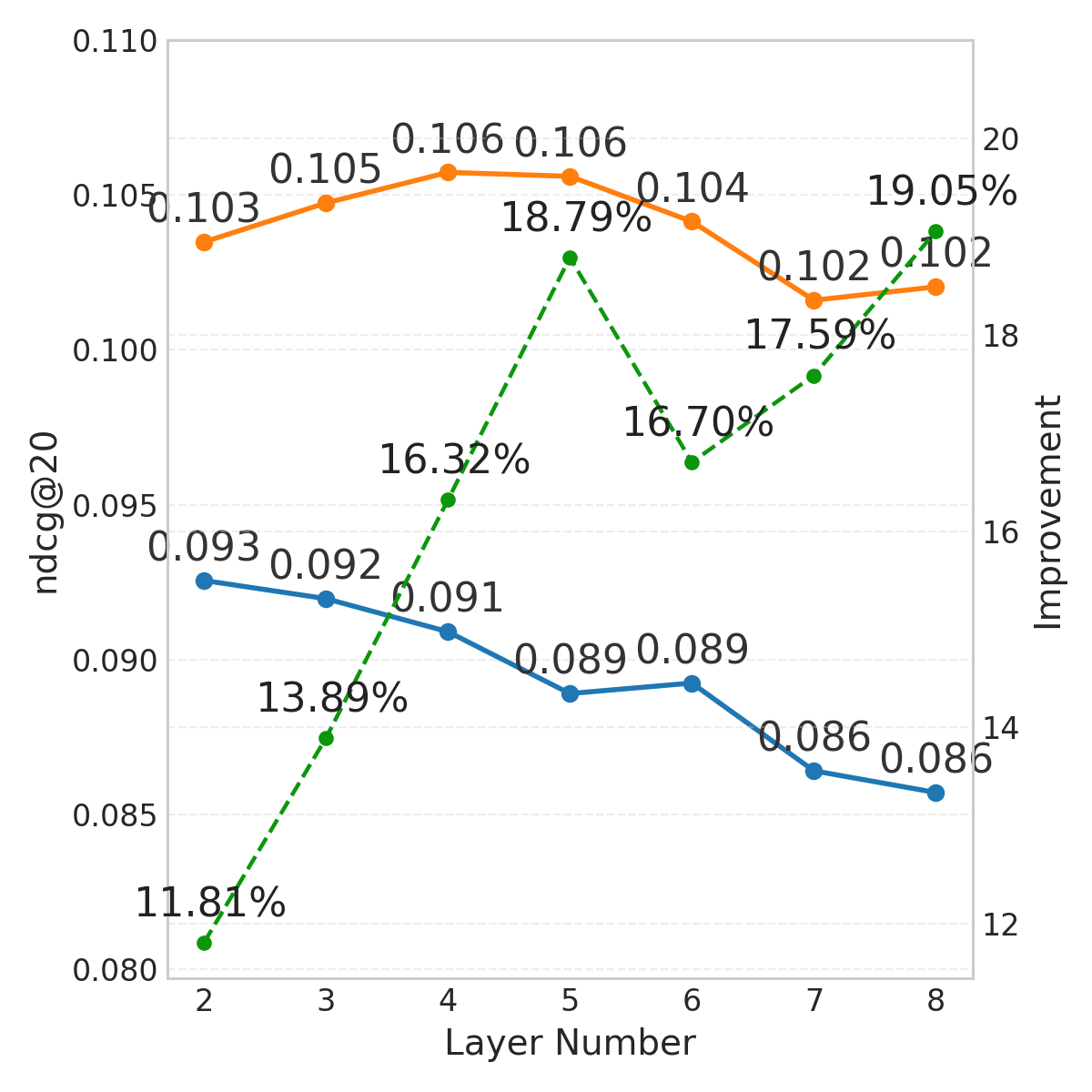}
        \label{fig:amazon_ndcg_20_vs_layer}
    }
    
    \vspace{0.01em} % 垂直间距
    
    % *** 第二行（两张图）***
    \subfloat[\fontsize{12pt}{13pt}\selectfont Yelp Recall@20 vs Layer Num.]{
        \includegraphics[width=0.48\textwidth]{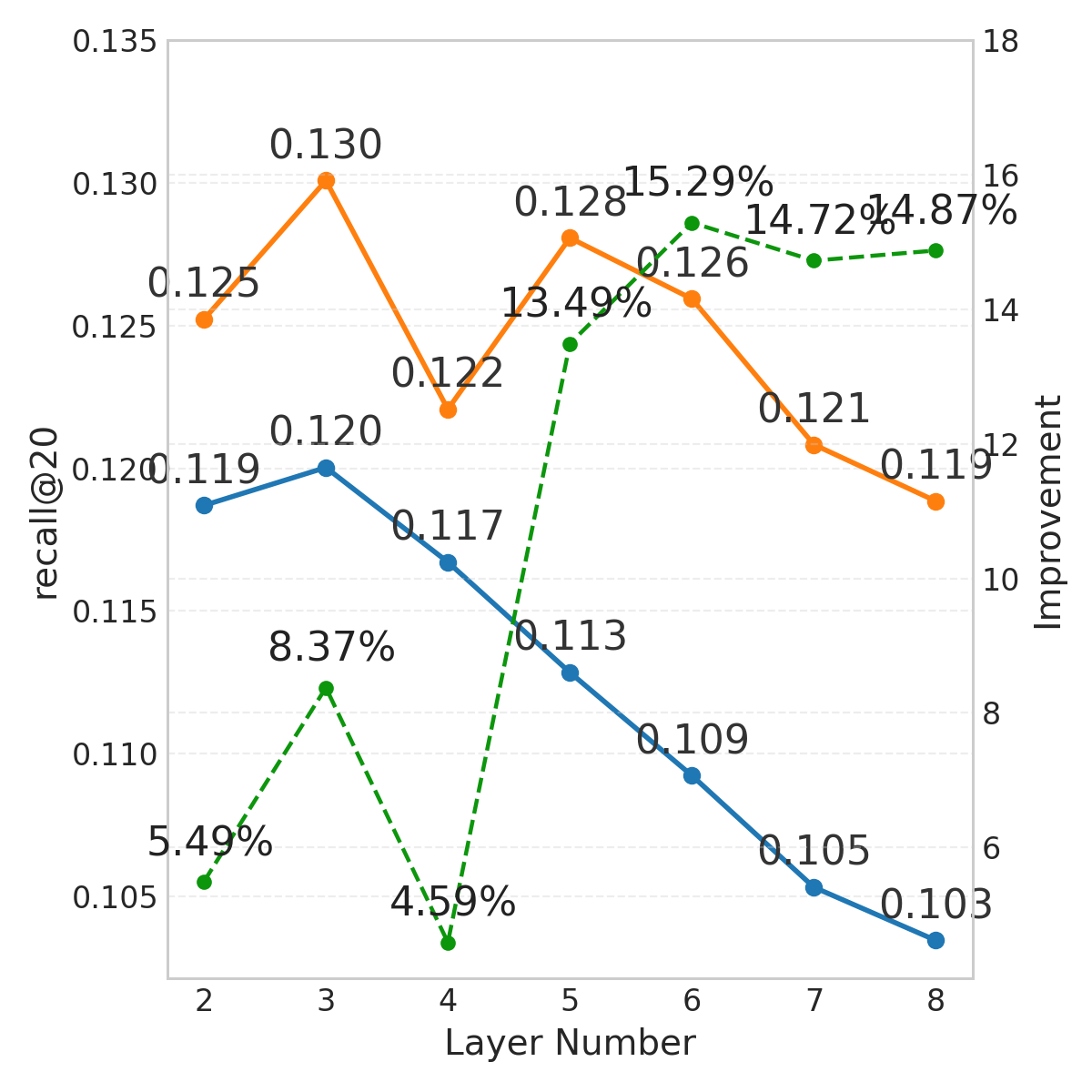}
        \label{fig:yelp_recall_20_vs_layer}
    }
    \hfill
    \subfloat[\fontsize{12pt}{13pt}\selectfont Yelp NDCG@20 vs Layer Num.]{
        \includegraphics[width=0.48\textwidth]{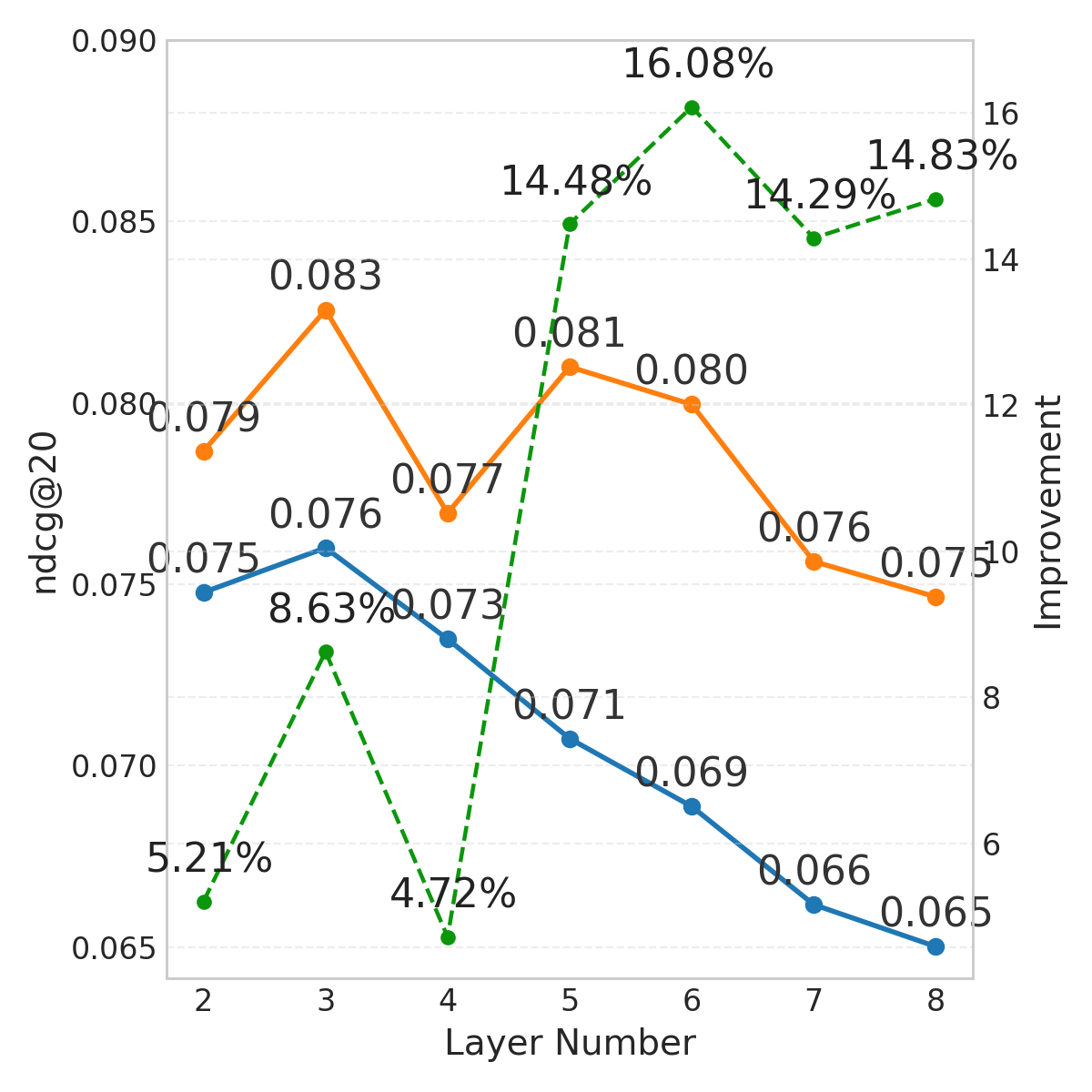}
        \label{fig:yelp_ndcg_20_vs_layer}
    }
}
    \vspace{0.1em} % 控制图例与上方子图的垂直距离
    \includegraphics[width=0.3\linewidth]{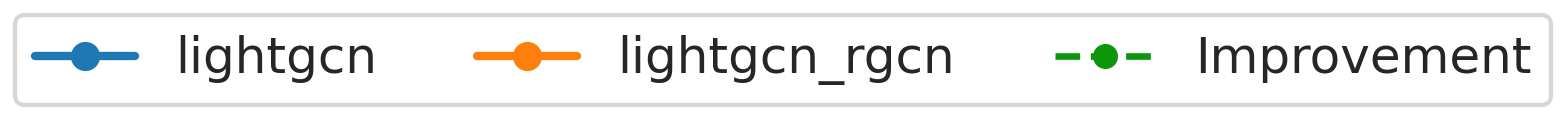} % 你的图例文件
    \vspace{-5mm}
    \caption{	Impact of layer count on TAGCF's  performances.}
    \label{fig:vs_layers}
\vspace{-2mm}
\end{figure*}
\textbf{Cold Start.}
To further examine the robustness of TAGCF under data-scarce conditions, we conduct a cold-start evaluation by simulating the introduction of unseen items. Specifically, we randomly hold out a certain proportion of items (controlled by the cold-start ratio $c$) along with all their corresponding interactions during training, and include them only in the testing phase.
For the baseline RLMRec, we allowed it to use pre-generated text embeddings for these items. For TAGCF, we dynamically generated attributes for these cold-start items by prompting LLMs with their metadata. The results, detailed in Table ~\ref{fig:cold_start_fig1} and ~\ref{fig:cold_start_fig2}, demonstrate that TAGCF maintains significantly higher performance than all baselines across different cold-start ratios.
Overall, these findings verify that TAGCF can effectively alleviate the cold-start issue by transforming implicit semantic cues into graph-structured knowledge that generalizes beyond observed interactions.

\begin{table}[t!]
\centering
\small
\setlength{\tabcolsep}{8pt}
\caption{Statistics of discovered paths.}
\vspace{-2mm}
\begin{tabular}{c|ccc}
\toprule
 & \textbf{Books} & \textbf{Yelp} & \textbf{Office} \\ 
\midrule
Test U-I Interaction & 40,106 & 55,436 & 23,289 \\
$\tau_{min}$ & 25 & 25 & 15 \\
$\tau_{max}$ & 5000 & 5000 & 5000 \\
Total 2-hop paths & 27,533,902 & 19,742,020 & 780,836 \\
Overlap Ratio & 0.4299 & 0.3543 & 0.0635 \\ 
\bottomrule
\end{tabular}
\label{tab:path_comparision}
\vspace{-3mm}
\end{table}

\subsection{Dive into the U-A-I graph}

\subsubsection{Attribute-Augmented Path} 
In the constructed U-A-I graph, a natural question arises: how many user-to-item paths mediated by attribute nodes are trustworthy, and how do they influence the effectiveness of TAGCF? To investigate this, we compute the composite adjacency matrix $\mathbf{G}_{Aug} =\mathbf{G}_{UA} \times \mathbf{G}^T_{VA}$ ,which captures paths starting from user nodes, passing through attribute nodes, and terminating at item nodes. While longer paths may exist in the graph, we focus on these length-2 paths for simplicity and interpretability.

We then compare $\mathbf{G}_{Aug}$ with 
$\mathbf{G}'_{UV}$, the adjacency matrix derived from the test set user-item interactions, to evaluate the overlap and coverage of discovered paths. The results are summarized in Table~\ref{tab:path_comparision}. In this table, Total 2-hop paths denotes the number of user–item connections mediated by attribute nodes, while Overlap Ratio measures the proportion of these 2-hop paths that coincide with real user–item interactions in the test set.
As shown in the table, both the Books and Yelp datasets achieve an overlap ratio exceeding 30\%, indicating that many discovered paths correspond to meaningful user–item relationships, which explains the substantial performance gains.
Even in the Office dataset, where the overlap ratio is relatively low, TAGCF still achieves around 10\% improvement across all base models, suggesting its robustness to noisy or sparse attribute connections.
These findings further demonstrate the feasibility and effectiveness of leveraging attributes as informative intermediate nodes in user–item graphs.

\vspace{-2.5mm}
 
\subsubsection{Anti-Over-Smoothing Analysis}
In this section, we demonstrate the {anti-over-smoothing ability} of TAGCF.
In traditional collaborative filtering–based recommendation methods, increasing the number of GNN layers does not always lead to better recommendation performance. On the contrary, deeper layers often degrade the results due to the over-smoothing phenomenon\cite{liu2020towards, chen2020measuring}, where node representations become indistinguishably similar. Some previous works attempt to mitigate this issue by constraining the correlations between node representations \cite{wu2024afdgcf, jin2022feature}.
Intuitively, this problem arises because convolution methods based on LightGCN ignore the importance weights between connected nodes. In LightGCN, each node’s representation can be transmitted to its neighbors without attenuation, even after multiple hops. However, introducing additional fully connected layers to after each convolution step, as done in NGCF, leads to a larger number of model parameters, which increases training difficulty and ultimately worsens recommendation performance.

TAGCF effectively alleviates this issue through its ARGC module. In ARGC, each convolution layer dynamically generates adaptive weights for individual nodes, thereby restricting lossless propagation between connected representations and preserving representation diversity across layers. Owing to the simple yet effective architecture of ARGC, only a limited number of additional parameters are introduced, preventing performance degradation. We conduct experiments on the Amazon-Book and Yelp datasets to investigate the over-smoothing phenomenon.
% Keeping all other hyperparameters fixed, we compare LightGCN and AGCF under different GNN layer depths. 
The experimental results are shown in Figure~\ref{fig:vs_layers}, where Improvement denotes the performance gain of TAGCF over LightGCN.
As shown in the figure, the improvement consistently increases as the number of layers grows. While the performance of LightGCN drops significantly with deeper layers, TAGCF maintains stable performance, demonstrating its strong ability to mitigate the over-smoothing problem.
\vspace{-2mm}

\subsection{Case Study}
\label{casestudy}

\begin{figure}
    \centering
    \includegraphics[width=\linewidth]{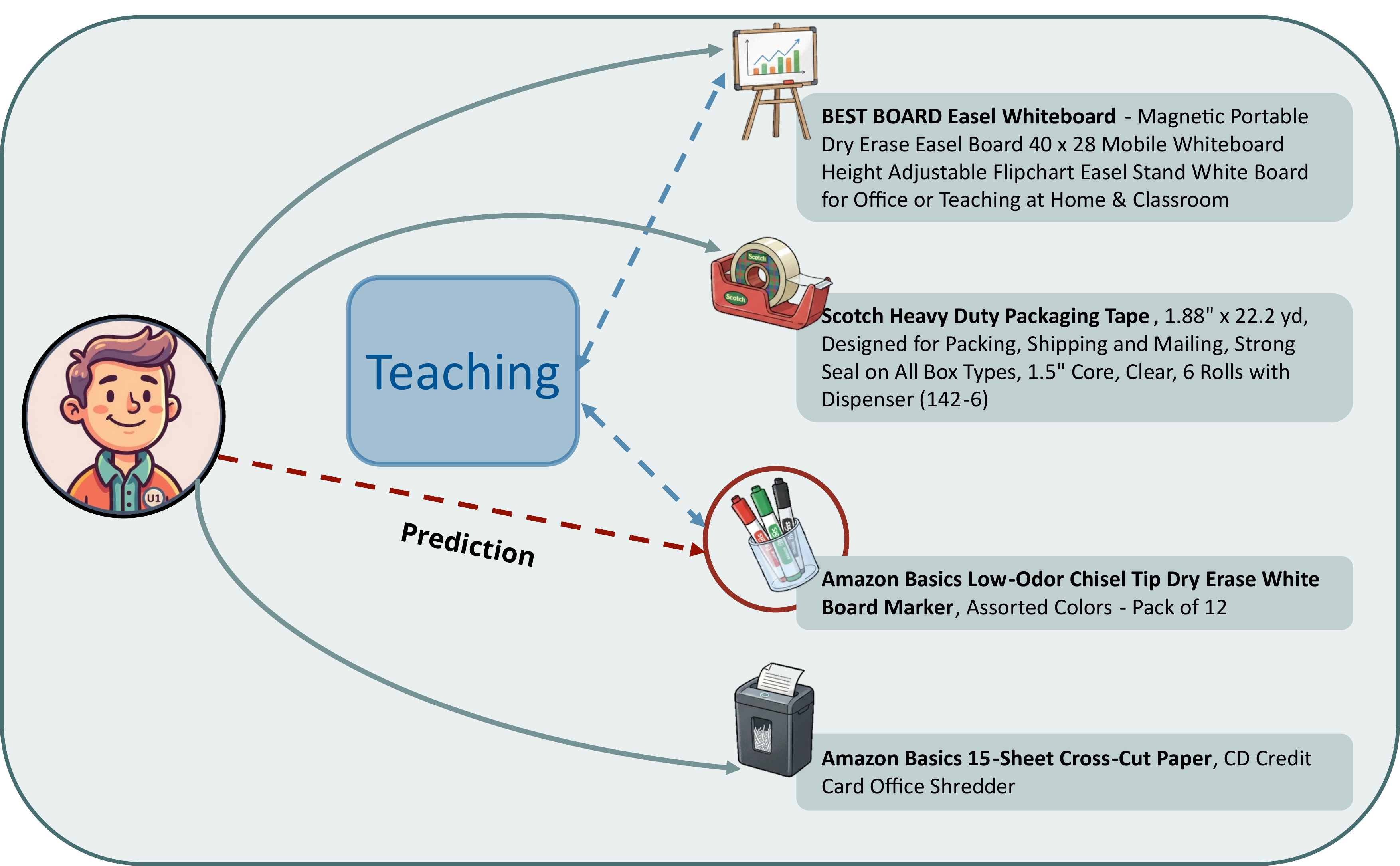}
    \vspace{-10pt}
    \caption{Case study on augmented path.}
    \label{fig:case_study}
    \vspace{-15pt}
\end{figure}

Here we provide a detailed case study from Office dataset to demonstrate the interpretability of our framework. We analyzed user ID \texttt{QAE35OPCE4E5LOQAODQT5QTD5SFIQ}, as illustrated in Figure~\ref{fig:case_study}. The training set for this user contains heterogeneous office supplies, including \textit{Best Board Easel Whiteboard}, \textit{Scotch Heavy Duty Packaging Tape}, and \textit{Amazon Basics 15-Sheet Cross-Cut Paper}.

While traditional collaborative filtering might view these simply as \textit{office items}, the LLM successfully extracted the explicit \textbf{Teaching} attribute from the \textit{Best Board Easel Whiteboard}. Crucially, this attribute bridges the semantic gap to a test-set item, \textit{Amazon Basics Low-Odor Chisel Tip Dry Erase White Board Marker}. This establishes a high-confidence causal pathway: User $\rightarrow$ Whiteboard $\rightarrow$ \textit{Teaching} (Intent) $\rightarrow$ Marker.

This pathway offers granular interpretability: it captures the \textit{complementary relationship} driven by the specific ``Teaching'' usage scenario, distinguishing it from the user's other interactions (e.g., the \textit{Shredder} or \textit{Tape}), which align with different latent intents like ``Logistic'' or ``Administration.'' Our ARGC module effectively captures this distinction by assigning a higher adaptive weight to the User-Attribute-Item subgraph associated with ``Teaching'', thereby amplifying the signal for the markers while suppressing noise from less relevant history. Consequently, the \textit{White Board Marker} rose from a 17th-place ranking to 2nd place. This case confirms that attribute extraction facilitates the identification of potential neighbor nodes for recently released or sparse items, effectively mitigating the cold-start problem through enhanced, interpretable representation in the embedding space.

\section{Conclusion}
% In this paper, we introduce a novel attribute-augmented framework, Attribute-Augmented Graph Collaborative Filtering(\textbf{AGCF}), designed for collaborative filtering recommender systems, along with its corresponding aggregation method Adaptive Relation-weighted Graph Convolution(\textbf{ARGC}). AGCF leverages large language models (LLMs) to generate potential hidden attributes inferred from user-item interactions. These inferred attributes are then used to construct a User-Attribute-Item (U-A-I) graph, where attributes are connected to their corresponding users and items. By incorporating these attributes, the user and item nodes are able to capture semantic signals derived from textual data through the aggregation process inherent in graph-based collaborative filtering, thereby enriching their original interaction-based representations.Given that different relationships contribute to the final embeddings of nodes in varying degrees, we propose ARGC, which employs a relation-specific network to predict the weights of each node in different layers in various graphs, and subsequently aggregates all representations to form the final embedding.
% In the evaluation, AGCF demonstrates state-of-the-art (SOTA) performance across all datasets and backbone architectures. Notably, AGCF is the only approach that surpasses methods using text embeddings without relying on large model-generated text embeddings. This highlights the effectiveness of AGCF and ARGC, and underscores the potential of causal reasoning through LLMs in the recommender system domain.

In this paper, we introduced TAGCF, a novel framework that leverages LLMs to infer interaction intents from user-item pairs, creating semantically meaningful attribute nodes in an enriched U-A-I graph. Unlike existing approaches that rely on textual embeddings, TAGCF enhances collaborative signals through structural graph enhancement.
To handle heterogeneous relations in the U-A-I graph, we proposed ARGC, which dynamically estimates relation importance and enables principled aggregation across user-attribute and attribute-item connections.
Extensive experiments demonstrate TAGCF's effectiveness across multiple CF backbones and datasets. Notably, TAGCF surpasses text-embedding methods without relying on textual features, achieving consistent state-of-the-art performance. Comprehensive evaluations validate that both attribute augmentation and adaptive relation weighting are essential for maximizing LLM-enhanced collaborative filtering benefits.

\newpage

\end{document}